\documentclass[pre,nofootinbib,superscriptaddress,twocolumn,floatfix,longbibliography]{revtex4-1}
\usepackage{graphicx}
\usepackage{amsmath}
\usepackage{amssymb}
\usepackage{tikz}
\usepackage{color} 
\usepackage{comment}
\usepackage{caption}
\usepackage{subcaption}
\newcommand{\rs}[1]{\textcolor{black}{#1}}

\usepackage{hyperref}
\hypersetup{colorlinks=true,linkcolor=blue,urlcolor=blue}

\begin{document}

\title{Defect Positioning in Combinatorial Metamaterials}

\author{Chaviva Sirote-Katz}
\email{chavivas@mail.tau.ac.il}
\affiliation{Department of Biomedical Engineering, Tel Aviv University, Tel Aviv 69978, Israel}

\author{Yotam M. Y. Feldman}
\affiliation{School of Chemistry, Tel Aviv University, Tel Aviv 69978, Israel}

\author{Guy Cohen}
\affiliation{School of Chemistry, Tel Aviv University, Tel Aviv 69978, Israel}
\affiliation{Center for Computational Molecular and Materials Science, Tel Aviv University, Tel Aviv 69978, Israel}

\author{Tam\'as K\'alm\'an}
\email{kalman@math.titech.ac.jp}
\affiliation{Department of Mathematics, Institute of Science Tokyo, H-214, 2-12-1 Ookayama, Meguro-ku, Tokyo 152-8551, Japan}
\affiliation{International Institute for Sustainability with Knotted Chiral Meta Matter (WPI-SKCM$^2$), Hiroshima University, Higashi-Hiroshima, Hiroshima 739-8526, Japan}

\author{Yair Shokef}
\email{shokef@tau.ac.il}
\affiliation{School of Mechanical Engineering, Tel Aviv University, Tel Aviv 69978, Israel}
\affiliation{School of Physics and Astronomy, Tel Aviv University, Tel Aviv 69978, Israel}
\affiliation{Center for Computational Molecular and Materials Science, Tel Aviv University, Tel Aviv 69978, Israel}
\affiliation{Center for Physics and Chemistry of Living Systems, Tel Aviv University, 69978, Tel Aviv, Israel}
\affiliation{International Institute for Sustainability with Knotted Chiral Meta Matter (WPI-SKCM$^2$), Hiroshima University, Higashi-Hiroshima, Hiroshima 739-8526, Japan}

\begin{abstract}

Combinatorial mechanical metamaterials are made of anisotropic, flexible blocks, such that multiple metamaterials may be constructed using a single block type, and the system's response depends on the frustration (or its absence) due to the mutual orientations of the blocks within the lattice. Specifically, any minimal loop of blocks that may not simultaneously deform in their softest mode defines a mechanical defect at the vertex (in two dimensions) or edge (in three dimensions) that the loop encircles. Defects stiffen the metamaterial, and allow to design the spatial patterns of stress and deformation as the system is externally loaded. We study the ability to place defects at arbitrary positions in metamaterials made of a family of block types that we recently introduced for the square, honeycomb, and cubic lattices. Alongside blocks for which we show that any defect configuration is possible, we identify situations in which not all sets are realizable as defects. One of the restrictions is that in three dimensions, defected edges form closed curves. Even in cases when not all geometries of defect lines are possible, we show how to produce defect lines of arbitrary knottedness. 

\end{abstract}

\maketitle

\section{Introduction} 
\label{sec:intro}

In mechanical metamaterials structures repeat in a lattice, such that the geometric deformation of each building block and the resultant interaction between neighboring blocks lead to non-trivial cooperative mechanical responses at the system level~\cite{bertoldi_flexible_review_2017, jang_soft_skin_2015, czajkowski_conformal_2022, guo_non-orientable_2023, chaco, kwakernaak_counting_2023}. Conventionally, and inspired by natural crystalline materials, most metamaterials are spatially periodic. However, since metamaterials are artificially designed and then fabricated at the macroscopic scale, one can construct not only any crystal- or quasi-crystal structure, but also arbitrary non-periodic structures. An approach to systematically consider such complex structures is based on anisotropic building blocks positioned on a regular lattice, where each repeating block may be independently oriented along any one of the principal directions of the lattice. The resulting class of metamaterials is referred to as \emph{combinatorial metamaterials}~\cite{coulais2016metacube}, since the different possible structures are defined by the discrete orientations of all blocks in the lattice, with the total number of possible metamaterials growing exponentially with the number of blocks in the lattice.

In a combinatorial metamaterial each building block typically has one soft mode of deformation, and the mutual orientations of neighboring blocks determine if they can all simultaneously deform according to their soft modes, thus forming a mechanically \emph{compatible} structure with a spatially extended soft mode of deformation. Alternatively, if neighboring blocks frustrate each other's ability to easily deform, the metamaterial lacks such a global soft mode and is thus more rigid and harder to deform. Frustration may be studied by considering minimal loops of blocks within the lattice; if the blocks surrounding a vertex (in two dimensions) or an edge (in three dimensions) may not deform in harmony, then that vertex or edge constitutes a \emph{mechanical defect}.

Geometric frustration occurs when the symmetry of the lattice is incompatible with the symmetry of the variables defined on it~\cite{Moessner2006}, with the classical example being the two-state antiferromagnetic Ising model on the three-fold rotationally symmetric triangular lattice~\cite{wannier1950}. Geometric frustration was classically studied in magnetic materials~\cite{bramwell2001}, but more recently also in artificially fabricated metamaterials, such as artificial spin ice~\cite{morrison2013, nisoli2017}, buckled colloidal monolayers~\cite{han2008, shokef2011, leoni2017}, colloidal ice~\cite{libal2006, ortiz-ambriz2016, oguz2020}, mechanical metamaterials~\cite{kang2014, udani2022} and arrays of rotors~\cite{mellado2012, teixeira2024}. In geometrically frustrated systems all minimal loops are frustrated, and the lattice is defected everywhere within it. However, frustration appears in non-periodic systems as well, most notably in spin glasses~\cite{Toulouse1977, Anderson1978, binder1986}, in which disorder causes some loops to be frustrated, while others are compatible. 

Combinatorial metamaterials allow to design \rs{(with certain limitations that we discuss below)} which loops will be compatible and which frustrated. Thus they are useful for exploring the effects of frustration and defects in physical systems in general~\cite{alexander2012}. Mechanical compatibility leads to combinatorial matching rules between neighboring units, with similar concepts employed in origami~\cite{dieleman_jigsaw_2020}. Manipulating the discrete orientations of the building blocks allows to design metamaterials with textured response~\cite{coulais2016metacube} or to use topological defects for spatially steering stress fields~\cite{meeussen2020supertriangles, pisanty2021putting}. Computational methods have been employed to obtain multiple desired responses of a given metamaterial when there are two soft modes per building block~\cite{bossart_oligomodal_2021, van_mastrigt_machine_2022, van_mastrigt_emergent_2023}.

In this paper we continue the investigation of a unified framework for generating combinatorial metamaterials in two and three dimensions, in which we consider all possible building blocks for the two-dimensional (2D) square and honeycomb lattices, as well as for the three-dimensional (3D) cubic lattice. An accompanying paper~\cite{compatible_paper} introduced this family of block types and studied the compatible metamaterials that may be constructed from them. Specifically, we examined the rate at which the number of compatible metamaterials grows with system size, and the ability to design the deformation texture on the boundary of the lattice. The present paper focuses on incompatible situations and explores whether one can design a metamaterial with an arbitrary spatial distribution of defects within it, meaning any assignment of which vertices or edges are defected and which are not.

In Sec.~\ref{sec:blocks} we systematically present all possible blocks. We refer interested readers to the accompanying paper~\cite{compatible_paper} for experimental realizations of all block types, as well as for a study of how the symmetries of some of them induce holographic order in compatible metamaterials. In the present work, for the square and honeycomb lattices, we introduce a duality relation that sorts the block types into pairs, such that the ability (or inability) to arbitrarily position defects in one block type implies the same for its dual pair, thus cutting by half the number of cases that need to be analyzed. In Sec.~\ref{sec:compatibility} we theoretically identify the conditions for mechanical (in)compatibility. We show that in 2D, defects are singular points, or vertices in the lattice, while in 3D, they are edges in the lattice that connect to form lines that cannot branch or terminate in the bulk, even though they can cross each other. 

In Sec.~\ref{sec:defects} we investigate the global structure of mechanical incompatibilities, or defects, where there is frustration between the desired deformations of neighboring blocks. We study the possible spatial distributions of point defects for the different 2D blocks, and the possible topological knotted or linked structures that defect lines can form for the various 3D blocks. Both in 2D and in 3D, we identify block types for which an arbitrary distribution of defects \rs{is not} realizable. However, in 3D we show that defect lines with arbitrary knottedness may be constructed using any one of the non-trivial blocks. Finally, in Sec.~\ref{sec:discussion} we discuss the results, and their consequences for further work on characterizing and designing physical responses of combinatorial metamaterials.

\section{Building Blocks} 
\label{sec:blocks}

\begin{figure*}[t!]
\centering
\includegraphics[width=\textwidth]{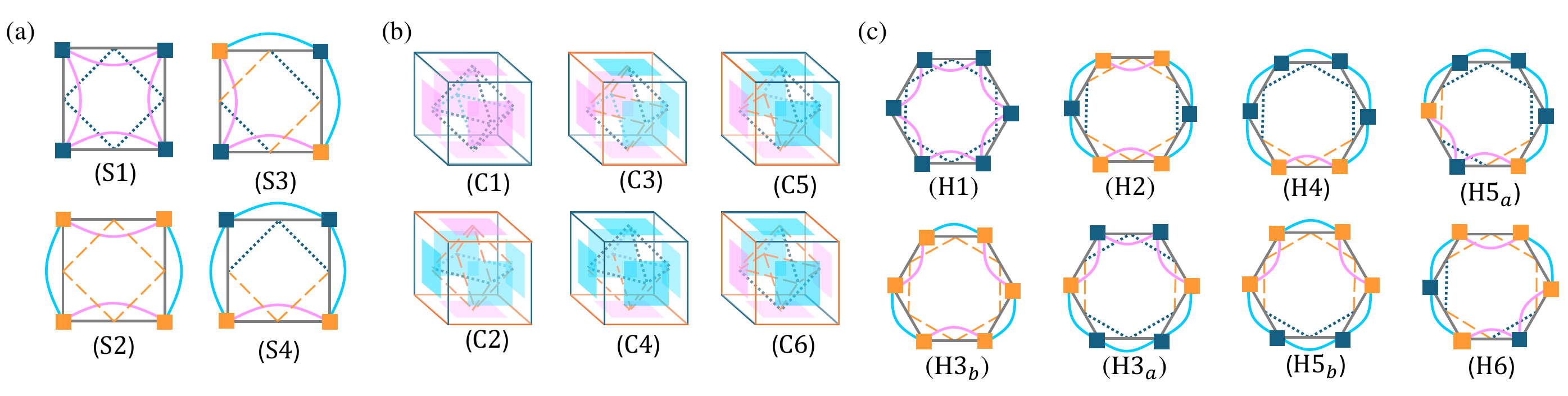}
\caption{\textbf{Building blocks.} All block types for the square (a), cubic (b), and honeycomb (c) lattices. For each of the possible blocks, we draw the struts (orange) and hinges (blue) that couple neighboring facets. In the 2D cases we mark the corresponding vertex with an orange or blue square, while in 3D we color the corresponding edge orange or blue. For all blocks we indicate one of the two polarizations of the soft mode of deformation. For the honeycomb lattice, we draw dual pairs one above the other.}
\label{fig:BB}
\end{figure*}

In this section, we briefly present the building blocks that were introduced, experimentally realized, and discussed in more detail in the accompanying paper~\cite{compatible_paper}. We introduce two additional ideas that will be useful for the analysis of defects in the metamaterial in the present paper: First, we associate hinges or struts between adjacent facets within each block to the vertex in 2D and edge in 3D that separates the two facets. Second, we identify a duality between block types, which will be especially useful in the analysis of realizability of defect configurations in the honeycomb lattice.

A given lattice structure implies a certain shape for its building blocks. Here, we restrict ourselves to blocks with a single soft mode, in which the motion of each of the block’s facets is either into or out of the block; all motions are orthogonal to the facet; and all have the same magnitude. Even with these restrictions, the soft mode of deformation of the block can take different discrete forms, which are characterized by whether each facet moves in or out with respect to the block. In this section we introduce a common framework for describing all possible blocks in different lattices, and specifically focus on the 2D square lattice, the 3D cubic lattice and the 2D honeycomb lattice.

\subsection{Square Lattice}

Perhaps the most widely studied mechanical metamaterial is based on the 2D square lattice, with a repeating structure that deforms most easily in the quadrupolar manner shown as Block~S2 in Fig.~\ref{fig:BB}a, see~\cite{resch_1965_patent, bertoldi_AdvMat2010, bar-sinai_charges_2020, Deng_domain_walls_2020, merrigan_PRR2021}. We extend this to consider all possible mechanical functionalities of a square building block. For a square block, each side can go in or out, leading to a total of $2^4=16$ possible deformations. However, deformation patterns that are related by inversion of all displacements correspond to the two possible \emph{polarizations} of a single block. In other words, these are the two directions in which the soft mode of the block may be actuated. Moreover, by rotational symmetry, the possible blocks reduce to the four shown in Fig.~\ref{fig:BB}a. 

In the drawings of Fig.~\ref{fig:BB}, in addition to schematically marking the two possible deformation directions of each side by a pink or blue arc, we draw a dashed orange line between adjacent sides such that when one contracts in to the block, the other protrudes out of it. This pair of sides may be thought of as being connected with a \emph{strut} that forces the two to move together. Conversely, when a pair of adjacent sides tends to both move in to the block or both out of it, we connect them with a dotted blue line, which represents a \emph{hinge}, where there is preference for bending the angle between the two sides, either to contract it or to expand it. Since a corner of the square block separates each pair of adjacent sides, we can equivalently mark the corners as possessing a strut (orange) or a hinge (blue). We will use these markings below when discussing compatibility, frustration, and defects. To simplify subsequent figures, later on we will omit the strut lines and will mark only the hinge lines. We can use the struts and hinges to recognize that Blocks~S1 and~S2 are isotropic, while Block~S3 has two possible orientations, and Block~S4 has four possible orientations. 

\subsection{Cubic Lattice}

We extend the above to 3D by considering the simple cubic lattice. As before, we assume the same magnitude of deformation along all six faces of the cubic building block, and use rotational symmetry to reduce the total number of $2^6 = 64$ displacement states to the six blocks shown in Fig.~\ref{fig:BB}b. Here as well, we denote by dashed orange lines the struts that connect adjacent faces so that when one moves into the block the other moves out of it; and we denote by dotted blue lines the hinges that connect adjacent faces such that both move in to the block or both move out of it. Here, each strut or hinge is associated with an edge of the cube that separates two adjacent faces, and we correspondingly color the edges of the blocks orange and blue for struts and hinges, respectively. We see that Block~C1 is isotropic, Block~C2 has three possible orientations, Block~C3 has four orientations, Blocks~C4 and C6 each have six possible orientations, and Block~C5 is the least symmetric, with $12$ possible orientations.

\subsection{Honeycomb Lattice}

The cubic lattice has three principal directions in three dimensions, and each cubic building block has three pairs of opposite facets. This is clearly related to the square lattice, which has two principal directions in two dimensions, and is made of square blocks, each with two pairs of opposite facets. However, there is another 2D lattice that we can compare to the 3D cubic lattice: the honeycomb lattice, which also has three principal directions, and its hexagonal building blocks have three pairs of opposite facets. Here too, we assume the same magnitude of deformation on all sides of the block, and use rotations and inversions to reduce the total number of $2^6 = 64$ displacement states to the eight blocks shown in Fig.~\ref{fig:BB}c.

\subsection{Duality}

The dashed orange lines denoting struts and the dotted blue lines denoting hinges in Fig.~\ref{fig:BB} assist in identifying how many distinct orientations each block type has. They also allow us to recognize a duality between blocks, whereby we say that a block with a certain arrangement of struts and hinges is dual to the one in which each strut is replaced by a hinge and each hinge is replaced by a strut. Specifically, Blocks~H1 and~H$3_b$ are dual and are both isotropic, Blocks~H2 and~H$3_a$ are dual and each has three possible orientations, finally Blocks~H4 and~H$5_b$ are dual and also Blocks~H$5_a$ and~H6 are dual, and each of these four may be oriented in six possible ways. 

By the same principle, in the 2D square lattice Blocks~S1 and~S2 are a dual pair, while Blocks~S3 and~S4 are each self dual. In the 3D cubic lattice such duality does not exist; each triplet of faces sharing a corner has three pairs of adjacent faces, which for consistency should be connected by an even number of struts. However, under the duality transformation, the parity of the number of struts connecting these three pairs of adjacent faces would become odd, which would not allow a consistent soft mode of deformation for the block.

\section{Mechanical Compatibility} 
\label{sec:compatibility}

For the sake of completeness, this section starts by repeating the explanation from the accompanying paper~\cite{compatible_paper} on the conditions for mechanical compatibility, whose failure to hold implies the presence of a mechanical defect. The second part of this section derives a constraint on the parity of defected edges meeting at a vertex in the 3D cubic lattice. This will be used in Sec.~\ref{sec:defects} when studying the ability to position defects in 3D.

\subsection{Local condition for compatibility}

For each of the aforementioned three lattices, we will consider combinatorial metamaterials constructed by positioning each block within the lattice in any one of its possible orientations. We will restrict ourselves to lattices in which all blocks are of the same type, although clearly many aspects of the present work may be generalized to lattices which mix different block types. To test for mechanical compatibility, we will follow loops of successive adjacent blocks within the lattice. Namely, we assume that one block deforms to one of its two polarizations, then based on the direction of deformation of its facets we deduce how its neighboring blocks should deform, and continue in such a sequence of blocks until we return to the block from which we started this analysis. If there is a self-consistent assignment of polarization for all blocks along this loop, we have mechanical compatibility, and at least as far as this loop is concerned, the blocks can all deform in tandem according to their soft mode. If, on the other hand, after completing the loop, we reach a contradiction, then we will say that this loop is frustrated, since the blocks along it cannot all simultaneously deform to their soft mode. 

We will test for compatibility locally. Namely, for each lattice, we will consider all the minimal loops within it. In 2D, a minimal loop is a loop containing the blocks surrounding a vertex in the lattice, and in 3D it contains the blocks surrounding an edge in the lattice. Any larger loop in the lattice may be constructed from a combination of such minimal loops, and therefore, if all the minimal loops are compatible, so is any larger loop that they form~\cite{meeussen2020supertriangles}. \rs{More precisely, let us emphasize that in this paper we only consider metamaterials enclosed in simply connected regions; in those cases, in the sense of mod $2$ chains, any loop can indeed be written as a sum of elementary loops. Furthermore, if we associate $0$ to compatible loops and $1$ to incompatible ones, we obtain a mod $2$ valued mapping that is additive on all loops.}

\begin{figure}[t]
\centering
\includegraphics[width=\columnwidth]{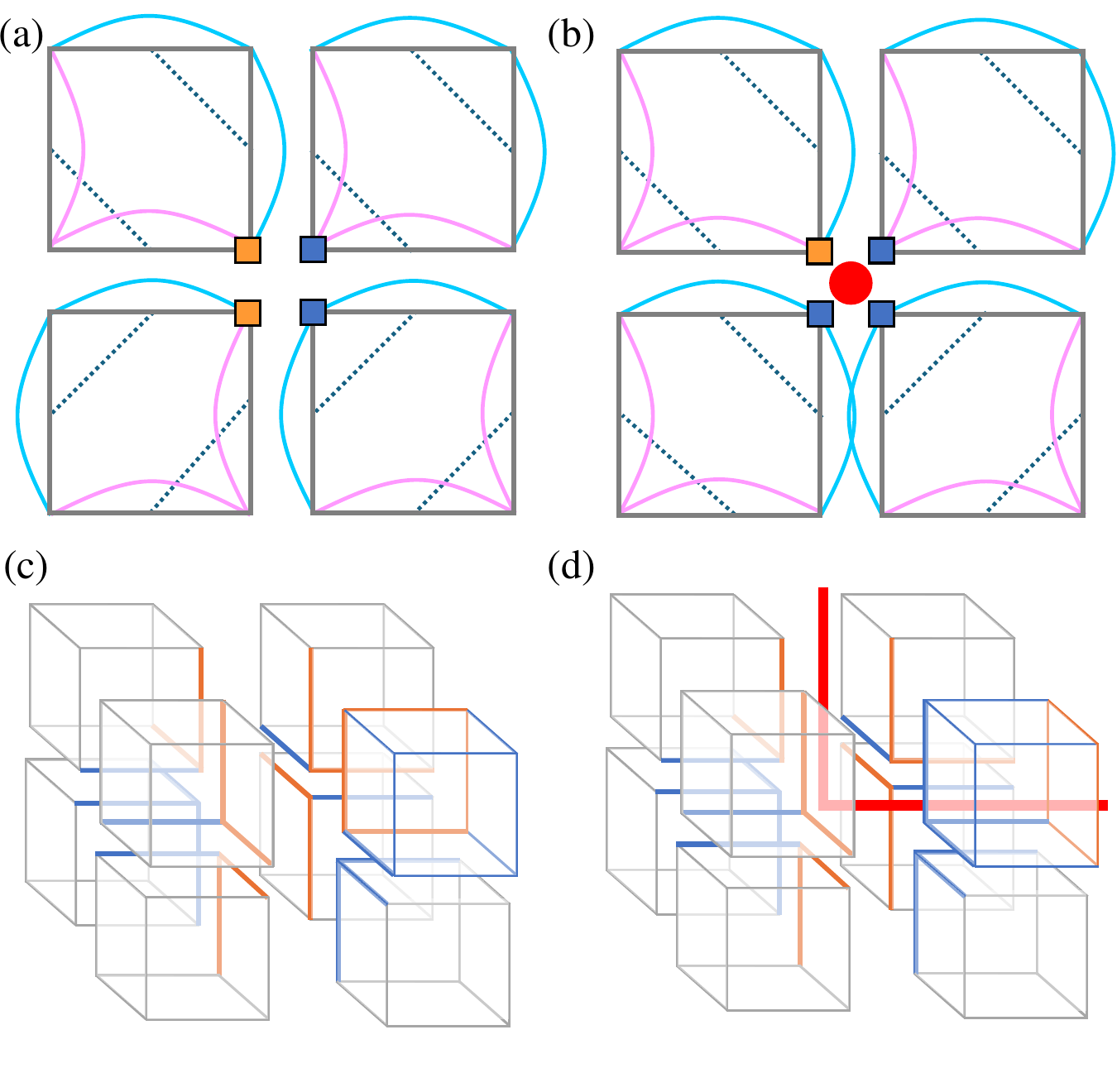}
\caption{\textbf{Mechanical compatibility.} Four Block~S3 square units can be mutually oriented, (a) in a compatible manner such that they can all simultaneously deform in their soft mode, or (b) in a frustrated manner in which there must be at least one block that cannot deform in its soft mode. The number of hinges incident to the central vertex is two (even) in (a) and three (odd) in (b). The red dot indicates the defected vertex at the center of the frustrated loop. \rs{(c) Eight Block~C4 cubic units oriented in a compatible manner; all six edges have an even number of hinges (blue lines). (d) A frustrated orientation of these eight units with two defected edges (red lines), at which there is an odd number of hinges.}}
\label{fig:compatibility}
\end{figure}

In Fig.~\ref{fig:compatibility}\rs{a,b} we examine mechanical compatibility for a minimal loop in the square lattice, which includes four building blocks around a vertex in the lattice. Here, we use Block~S3 to demonstrate that certain orientations of the blocks give a compatible structure, while other orientations lead to frustration. If a minimal loop in 2D is frustrated, we can identify the vertex that this loop encircles as a mechanical defect in the lattice. In 3D, edges will be the basic objects that are either compatible or frustrated (a.k.a.\ defected). 

\rs{In 2D, we use the following formalism to determine if a vertex is defected or not. We start with an arbitrary facet surrounding this vertex and 
associate to it $S_0=1$ to indicate that it wishes to deform clockwise around the vertex. As we move to the next adjacent facet, if the corner of the block separating the two facets includes a strut, then that next facet will prefer to deform in the same orientation, that is $S_1=S_0=1$, while if the block contains a hinge, the direction of rotation around the vertex will flip, resulting in $S_1=-S_0=-1$. As we encircle the vertex and return to the initial facet, a compatible deformation of all blocks will, therefore, require this facet to deform in the rotational direction described by $S=(-1)^h$, where $h$ is the number of hinges incident to the vertex. Clearly, if the number of hinges is even, this is equal to $S_0=1$ and the blocks around this vertex can simultaneously deform to their soft mode, while if $h$ is odd, the vertex is frustrated and we identify it as a mechanical defect. A similar analysis can be carried out in 3D, where we consider the blocks surrounding an edge of the lattice, and conclude that the defectedness of that edge depends on the parity of hinges around it.}

\subsection{Defects form closed curves}

In 3D, there is an important local restriction on what sets of nearby edges may be simultaneously defected. Namely, we will now show that in any 3D metamaterial, even if we mix all six types of blocks together and use them with arbitrary orientations, the frustrated edges form a graph in which the degree of each interior node is even. Therefore, defect lines can never end in the bulk of the material because such an endpoint would have degree one. Rather, defect lines either end at the boundary of the metamaterial, or form closed curves within the bulk. Note that degree four and degree six are valid options, that is, defect lines may intersect, even in triple points, however they may not branch, which would require three (or five) defect lines to meet at a vertex.

First, consider a single cubic block. Its 2D surface is divided into regions of two different colors, indicating inward and outward motion in the soft mode, as depicted in Fig.~\ref{fig:BB}b. \rs{Adjacent faces of the same color are separated by a hinge, and a pair of opposite colors are separated by a strut.} The one-dimensional boundary of each color is a cycle, i.e., it has even degree at each vertex. Since two adjacent faces connected by a hinge have the same color and two faces connected with a strut have opposite colors, it follows that there is an even number of struts at each vertex. Now, since each vertex of the cubic block has total degree three, the number of hinges at each vertex is odd.

When considering the cubic lattice, each edge is shared by four blocks, \rs{as shown in Fig.~\ref{fig:compatibility}c,d}. Each of these four blocks has either a strut or a hinge at that edge, and the edge is defected if the number of hinges is odd. The parity of defected edges meeting at a vertex of the lattice is the parity of the total number of these hinges. This number may be considered by summing over the six edges, where for each edge we sum over the four blocks surrounding it. Alternatively, we can sum over the eight blocks around the vertex, noting that each such block has three edges, where as discussed above, the number of hinges among the three must be odd. Since the sum of eight odd numbers is even, the number of defected edges meeting at the vertex must be even.

\section{Defect Positioning}
\label{sec:defects}

In this section we analyze the possible defect sets that may occur in metamaterials constructed from each one of the building blocks reviewed in Sec.~\ref{sec:blocks}. We will ask, for the different block types, whether one can design metamaterials that will have defects in any desired spatial distribution. As we have just discussed, in 3D, only configurations with an even number of defected edges meeting at every vertex of the lattice are feasible, and we will thus consider only such cases.

The problem at hand is rather similar in nature to the boundary texture design problem treated in~\cite{compatible_paper}. There we wanted no defects in the interior, which is a special case of defect positioning, whereas along the boundary, at each vertex (in 2D cases) or edge (in 3D cases), the parity of the number of hinges was prescribed by the desired texture -- just like in the defect positioning problem. The difference is, in a sense, topological: instead of having to work inward from a boundary condition, this time we can start in the interior and work outward.

As explained in~\cite{compatible_paper}, metamaterials made of the isotropic Blocks~S1, S2, H$3_b$, or C1 do not have any defects in them, while in the metamaterial made of Block~H1 all vertices are defected. In what follows, for Blocks~S3, S4, H4, H$5_a$, H$5_b$, H6, C3, C4, C5, and C6 we provide schemes for constructing any defect configuration. On the other hand for Blocks~H2, H$3_a$, and C2 we present both counting arguments and counterexamples showing that not all defect configurations may be realized. 

Additionally, especially for cubic Block~C2, one can also investigate the following topological question: Can the closed curves, formed by the defected edges in the metamaterial, realize any given knot or link type? Since any knot or link can be realized by paths in the lattice, the answer is obviously yes for Blocks~C3, C4, C5, and C6. In the remaining non-trivial case of Block~C2, this is not so automatic but we will show how to obtain any topological knotting and linking of the defect line. 

\subsection{Square Lattice}

\begin{figure}[t]
\centering
\includegraphics[width=.8\columnwidth]{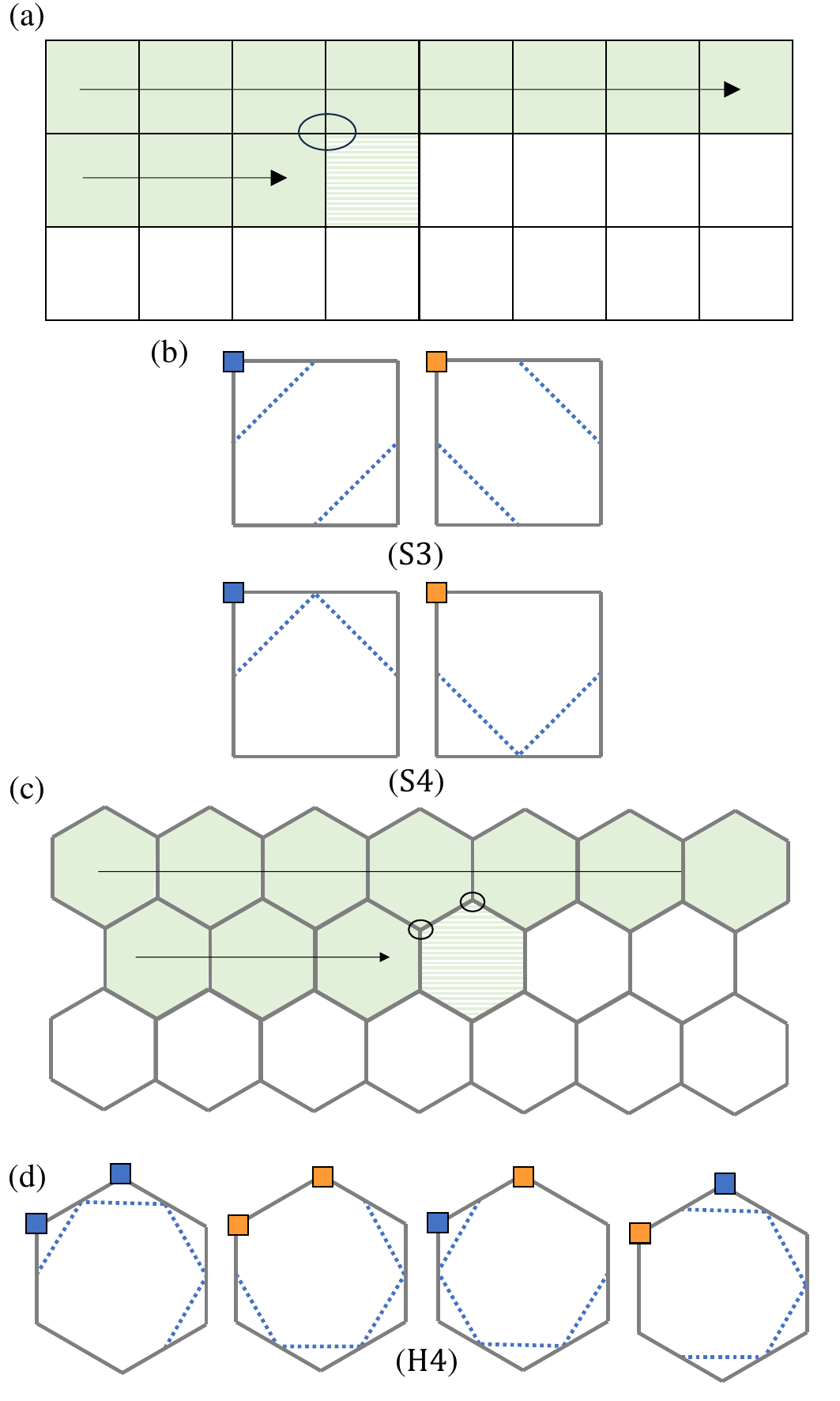}
\caption{\textbf{Scanning the square and honeycomb lattices to position defects.} The orientations of the green blocks have already been assigned, in the square (a) and honeycomb (c) lattices. Adding the new, hashed block completes a loop of blocks around one vertex in the square lattice (circled in (a)) and around two vertices in the honeycomb lattice (circled in (c)). Each of these vertices can be needy (blue) or not (orange). (b) We show orientations of Blocks~S3, and~S4 that meet each requirement. With Block~H4, for each of the four possible neediness situations, we show in (d) an orientation that provides a fit.}
\label{fig:anydefect}
\end{figure}

Blocks~S1 and S2 are isotropic, and metamaterials built from them do not have any defects. With Blocks~S3 and S4, on the other hand, we may realize any defect distribution, in other words any desired assignment of which vertices in the lattice are defected and which are not. This can be established by a basic scanning argument outlined in Fig.~\ref{fig:anydefect}a,b. As we scan the lattice, we will ask whether each vertex that we enclose needs there to be a hinge in the last block around it, or not. More generally, for Blocks~S3 and S4, as well as in the analysis that will follow for the honeycomb lattice, we say that a vertex is \emph{needy}, at some stage in the construction of the metamaterial, if the parity of the number of surrounding hinges, in the blocks that have already been set, is the opposite of what is required for the desired defectedness status of the vertex. In particular, with regard to the last block to be positioned next to a vertex, a needy vertex requires a hinge in this last block, while a \emph{non-needy} vertex requires the absence of a hinge, that is, a strut. Since both Block~S3 and Block~S4 can be oriented either with or without contributing a hinge to this vertex, we can always continue this scanning process to eventually design a metamaterial with any desired distribution of defects in the lattice.

\subsection{Honeycomb Lattice}

Here the notion of duality, which we introduced at the end of Sec.~\ref{sec:blocks}, will prove to be helpful. Recall that hexagonal blocks form the dual pairs (H1,H3$_b$), (H2,H3$_a$), (H4,H5$_b$), and (H5$_a$,H6), where the duality transformation switches between hinges and struts. Each vertex in the honeycomb lattice is surrounded by three hexagons, and each of them can contribute either a hinge or a strut to the loop surrounding that vertex. Thus, whenever a vertex is defected in a metamaterial built from a certain block type, the same vertex is non-defected when all blocks are replaced by their duals, and vice versa, as the parity of hinges in the loop around it switches.

For instance, in the metamaterial made of Block~H1, all vertices are defected, while the metamaterial made of its dual Block~H3$_b$ has no defects at all~\cite{compatible_paper}. For the remaining dual block pairs, we analyze one member of each pair and then from the result we deduce our conclusion for the dual block, too. Specifically, if there are defect configurations that may not be realized with a certain block type, then the dual defect configuration, with defected and non-defected vertices interchanged, is not realizable with the dual block. Similarly, if all defect configurations are realizable with a certain block type, they are also realizable by the dual block.

\subsubsection{Blocks~H2 and H3$_a$}

\begin{figure}[t]
\begin{tikzpicture}[scale=.15]
\node at (-10, 9) {\small (a)};
\draw[shift={(0,10.26)}] (-1,-1.71)--(0,0)--(2,0)--(3,-1.71);
\draw[shift={(3,8.55)}] (-1,-1.71)--(0,0)--(2,0)--(3,-1.71);
\draw[shift={(6,6.84)}] (-1,-1.71)--(0,0)--(2,0)--(3,-1.71);
\draw[shift={(9,5.13)}] (-1,-1.71)--(0,0)--(2,0)--(3,-1.71);
\draw[shift={(-3,8.55)}] (-1,-1.71)--(0,0)--(2,0)--(3,-1.71);
\draw[shift={(0,6.84)}] (-1,-1.71)--(0,0)--(2,0)--(3,-1.71);
\draw[shift={(3,5.13)}] (-1,-1.71)--(0,0)--(2,0)--(3,-1.71);
\draw[shift={(6,3.42)}] (-1,-1.71)--(0,0)--(2,0)--(3,-1.71);
\draw[shift={(9,1.71)}] (-1,-1.71)--(0,0)--(2,0)--(3,-1.71);
\draw[shift={(-6,6.84)}] (-1,-1.71)--(0,0)--(2,0)--(3,-1.71);
\draw[shift={(-3,5.13)}] (-1,-1.71)--(0,0)--(2,0)--(3,-1.71);
\draw[shift={(0,3.42)}] (-1,-1.71)--(0,0)--(2,0)--(3,-1.71);
\draw[shift={(3,1.71)}] (-1,-1.71)--(0,0)--(2,0)--(3,-1.71);
\draw[shift={(6,0)}] (-1,-1.71)--(0,0)--(2,0)--(3,-1.71);
\draw[shift={(9,-1.71)}] (-1,-1.71)--(0,0)--(2,0)--(3,-1.71);
\draw[shift={(-6,3.42)}] (-1,-1.71)--(0,0)--(2,0)--(3,-1.71);
\draw[shift={(-3,1.71)}] (-1,-1.71)--(0,0)--(2,0)--(3,-1.71);
\draw (-1,-1.71)--(0,0)--(2,0)--(3,-1.71);
\draw[shift={(3,-1.71)}] (-1,-1.71)--(0,0)--(2,0)--(3,-1.71);
\draw[shift={(6,-3.42)}] (-1,-1.71)--(0,0)--(2,0)--(3,-1.71);
\draw[shift={(-6,0)}] (-1,-1.71)--(0,0)--(2,0)--(3,-1.71);
\draw[shift={(-3,-1.71)}] (-1,-1.71)--(0,0)--(2,0)--(3,-1.71);
\draw[shift={(0,-3.42)}] (-1,-1.71)--(0,0)--(2,0)--(3,-1.71);
\draw[shift={(3,-5.13)}] (-1,-1.71)--(0,0)--(2,0)--(3,-1.71);
\draw[shift={(-6,-3.42)}] (-1,-1.71)--(0,0)--(2,0)--(3,-1.71);
\draw[shift={(-3,-5.13)}] (-1,-1.71)--(0,0)--(2,0)--(3,-1.71);
\draw[shift={(0,-6.84)}] (-1,-1.71)--(0,0)--(2,0)--(3,-1.71);
\draw (-7,5.13)--(-6,3.42);
\draw (-7,1.71)--(-6,0);
\draw (-7,-1.71)--(-6,-3.42);
\draw (-7,-5.13)--(-6,-6.84)--(-4,-6.84)--(-3,-8.55)--(-1,-8.55)--(0,-10.26)--(2,-10.26)--(3,-8.55)--(5,-8.55)--(6,-6.84)--(8,-6.84)--(9,-5.13)--(11,-5.13)--(12,-3.42);
\draw (11,-1.71)--(12,0);
\draw (11,1.71)--(12,3.42);
\draw[radius=.5,red,fill] (-1,5.13) circle;
\draw[radius=.5,red,fill] (5,5.13) circle;
\draw[radius=.5,red,fill] (-4,0) circle;
\draw[radius=.5,red,fill] (2,0) circle;
\draw[radius=.5,red,fill] (8,0) circle;
\draw[radius=.5,red,fill] (-1,-5.13) circle;
\draw[radius=.5,red,fill] (5,-5.13) circle;
\begin{scope}[shift={(30,0)}]
\node at (-10, 9) {\small (b)};
\draw[shift={(0,10.26)}] (-1,-1.71)--(0,0)--(2,0)--(3,-1.71);
\draw[shift={(3,8.55)}] (-1,-1.71)--(0,0)--(2,0)--(3,-1.71);
\draw[shift={(6,6.84)}] (-1,-1.71)--(0,0)--(2,0)--(3,-1.71);
\draw[shift={(9,5.13)}] (-1,-1.71)--(0,0)--(2,0)--(3,-1.71);
\draw[shift={(-3,8.55)}] (-1,-1.71)--(0,0)--(2,0)--(3,-1.71);
\draw[shift={(0,6.84)}] (-1,-1.71)--(0,0)--(2,0)--(3,-1.71);
\draw[shift={(3,5.13)}] (-1,-1.71)--(0,0)--(2,0)--(3,-1.71);
\draw[shift={(6,3.42)}] (-1,-1.71)--(0,0)--(2,0)--(3,-1.71);
\draw[shift={(9,1.71)}] (-1,-1.71)--(0,0)--(2,0)--(3,-1.71);
\draw[shift={(-6,6.84)}] (-1,-1.71)--(0,0)--(2,0)--(3,-1.71);
\draw[shift={(-3,5.13)}] (-1,-1.71)--(0,0)--(2,0)--(3,-1.71);
\draw[shift={(0,3.42)}] (-1,-1.71)--(0,0)--(2,0)--(3,-1.71);
\draw[shift={(3,1.71)}] (-1,-1.71)--(0,0)--(2,0)--(3,-1.71);
\draw[shift={(6,0)}] (-1,-1.71)--(0,0)--(2,0)--(3,-1.71);
\draw[shift={(9,-1.71)}] (-1,-1.71)--(0,0)--(2,0)--(3,-1.71);
\draw[shift={(-6,3.42)}] (-1,-1.71)--(0,0)--(2,0)--(3,-1.71);
\draw[shift={(-3,1.71)}] (-1,-1.71)--(0,0)--(2,0)--(3,-1.71);
\draw (-1,-1.71)--(0,0)--(2,0)--(3,-1.71);
\draw[shift={(3,-1.71)}] (-1,-1.71)--(0,0)--(2,0)--(3,-1.71);
\draw[shift={(6,-3.42)}] (-1,-1.71)--(0,0)--(2,0)--(3,-1.71);
\draw[shift={(-6,0)}] (-1,-1.71)--(0,0)--(2,0)--(3,-1.71);
\draw[shift={(-3,-1.71)}] (-1,-1.71)--(0,0)--(2,0)--(3,-1.71);
\draw[shift={(0,-3.42)}] (-1,-1.71)--(0,0)--(2,0)--(3,-1.71);
\draw[shift={(3,-5.13)}] (-1,-1.71)--(0,0)--(2,0)--(3,-1.71);
\draw[shift={(-6,-3.42)}] (-1,-1.71)--(0,0)--(2,0)--(3,-1.71);
\draw[shift={(-3,-5.13)}] (-1,-1.71)--(0,0)--(2,0)--(3,-1.71);
\draw[shift={(0,-6.84)}] (-1,-1.71)--(0,0)--(2,0)--(3,-1.71);
\draw (-7,5.13)--(-6,3.42);
\draw (-7,1.71)--(-6,0);
\draw (-7,-1.71)--(-6,-3.42);
\draw (-7,-5.13)--(-6,-6.84)--(-4,-6.84)--(-3,-8.55)--(-1,-8.55)--(0,-10.26)--(2,-10.26)--(3,-8.55)--(5,-8.55)--(6,-6.84)--(8,-6.84)--(9,-5.13)--(11,-5.13)--(12,-3.42);
\draw (11,-1.71)--(12,0);
\draw (11,1.71)--(12,3.42);
\draw[radius=.5,red,fill] (0,6.84) circle;
\draw[radius=.5,red,fill] (2,6.84) circle;
\draw[radius=.5,red,fill] (-3,5.13) circle;
\draw[radius=.5,red,fill] (3,5.13) circle;
\draw[radius=.5,red,fill] (-4,3.42) circle;
\draw[radius=.5,red,fill] (0,3.42) circle;
\draw[radius=.5,red,fill] (2,3.42) circle;
\draw[radius=.5,red,fill] (6,3.42) circle;
\draw[radius=.5,red,fill] (8,3.42) circle;
\draw[radius=.5,red,fill] (-3,1.71) circle;
\draw[radius=.5,red,fill] (-1,1.71) circle;
\draw[radius=.5,red,fill] (3,1.71) circle;
\draw[radius=.5,red,fill] (5,1.71) circle;
\draw[radius=.5,red,fill] (9,1.71) circle;
\draw[radius=.5,red,fill] (0,0) circle;
\draw[radius=.5,red,fill] (6,0) circle;
\draw[radius=.5,red,fill] (-3,-1.71) circle;
\draw[radius=.5,red,fill] (-1,-1.71) circle;
\draw[radius=.5,red,fill] (3,-1.71) circle;
\draw[radius=.5,red,fill] (5,-1.71) circle;
\draw[radius=.5,red,fill] (9,-1.71) circle;
\draw[radius=.5,red,fill] (-4,-3.42) circle;
\draw[radius=.5,red,fill] (0,-3.42) circle;
\draw[radius=.5,red,fill] (2,-3.42) circle;
\draw[radius=.5,red,fill] (6,-3.42) circle;
\draw[radius=.5,red,fill] (8,-3.42) circle;
\draw[radius=.5,red,fill] (-3,-5.13) circle;
\draw[radius=.5,red,fill] (3,-5.13) circle;
\draw[radius=.5,red,fill] (0,-6.84) circle;
\draw[radius=.5,red,fill] (2,-6.84) circle;
\end{scope}
\end{tikzpicture}
\caption{\textbf{Non-realizable defect configurations in the honeycomb lattice.} Complementer defect configurations that are non-realizable for Block H3$_a$~(a) and its dual Block~H2~(b).}
\label{fig:complementers}
\end{figure}
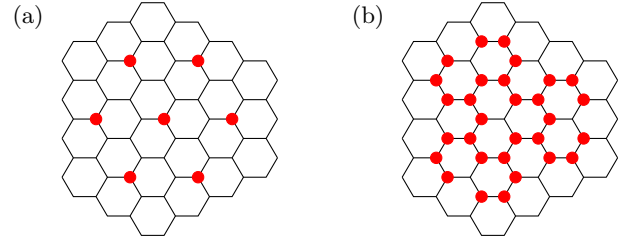

For Block~H3$_a$, Fig.~\ref{fig:complementers}a shows a configuration of seven vertices that cannot be defected with this block type, provided that all other vertices are non-defected. The symmetry of the configuration makes it easy to check this claim by separating a few cases. The dual set, shown in Fig.~\ref{fig:complementers}b, in which all vertices are defected except for those seven, cannot be realized by Block~H2. 

The existence of such examples follows from a simple counting argument, too. A hexagonal region of side $L$ contains $B=3L^2-3L+1$ hexagonal blocks. Since each of these blocks has three possible orientations, the total number of possible metamaterials of this size is $M=3^{B}$. On the other hand, the number of internal vertices within this region is $V=6(L-1)^2$. Since a defect configuration specifies whether each one of these vertices is defected or not, there could be $D=2^{V}$ possible defect configurations. However, for $L \ge 6$ we have $D>M$, that is, there are more possible defect configurations than metamaterials of this size. Therefore, for large enough lattices there are defect configurations that cannot be realized. This may also be seen asymptotically, noting that for $L \gg 1$ we have $M \propto 27^L$, while $D \propto 64^L$. In particular, the proportion of realizable defect sets tends exponentially to zero as $L$ approaches infinity. Note however that the example shown in Fig.~\ref{fig:complementers} fits inside a hexagon of side length $L=4$. This demonstrates that already for systems smaller than the threshold of $L=6$, some defect configurations may not be realized. Either way, for Blocks~H2 and H3$_a$, we see that not every defect configuration is possible; however beyond that, we are not aware of a simple criterion for realizability of a distribution of defects using these block types. 

The remaining hexagonal Blocks~H4, H5$_a$, H5$_b$, and H6 each have six distinct orientations, so the above counting argument does not rule out the realizability of any defect distribution for these blocks. However, it does not imply that any defect distribution is realizable, either. In what follows we provide a detailed analysis for these four block types and show that with all four, any defect configuration is in fact feasible.

\subsubsection{Blocks~H4 and H5$_b$}

With Blocks H4 and H5$_b$, the fact that any defect configuration is realizable can be shown similarly to what we did above for the square lattice. If we simply scan line-by-line, then in the honeycomb lattice, each added block closes loops around two new vertices (Fig.~\ref{fig:anydefect}c), and not only around one vertex as in the square lattice (Fig.~\ref{fig:anydefect}a). However, as shown in Fig.~\ref{fig:anydefect}d, Block~H4 may be oriented so as to add a hinge to the loops around any one of these two adjacent vertices, or to both of them, or to none of them. This allows to satisfy the neediness of all vertices in the lattice, and by that to obtain any desired spatial distribution of defects. This same result follows by duality for Block~H5$_b$.

\subsubsection{Blocks~H5$_a$ and H6}

Also for the last two hexagonal units, Blocks~H5$_a$ and~H6, arbitrary defect sets are possible. By duality it suffices to establish this for one of them, and in Appendix~\ref{sec:caterpillar} we give a construction protocol for Block~H6. It is still based on the principle of scanning the lattice, but this time a more careful implementation is necessary due to the fact that Block~H6 does not have a pair of adjacent hinges, and thus cannot always be oriented to satisfy the neediness of two adjacent vertices. Namely, the first of the four options shown for Block~H4 in Fig.~\ref{fig:anydefect}d should be avoided when scanning the lattice with Block~H6. 

\subsection{Cubic Lattice}

The isotropic Block~C1 can only form a single, trivial metamaterial with no defected edges, and it will be excluded from the rest of the discussion. Metamaterials made of the other block types can have defected edges that form lines within the lattice. As we have shown in Sec.~\ref{sec:compatibility} above, the number of defected edges meeting at a vertex must be even. That is, the set of defected edges obeys a local restriction that defect lines cannot terminate or branch in the bulk of the metamaterial.

\subsubsection{Blocks~C3, C4, C5 defect realization by scanning}

\begin{figure}[t!]
\centering
\includegraphics[width=.7\columnwidth]{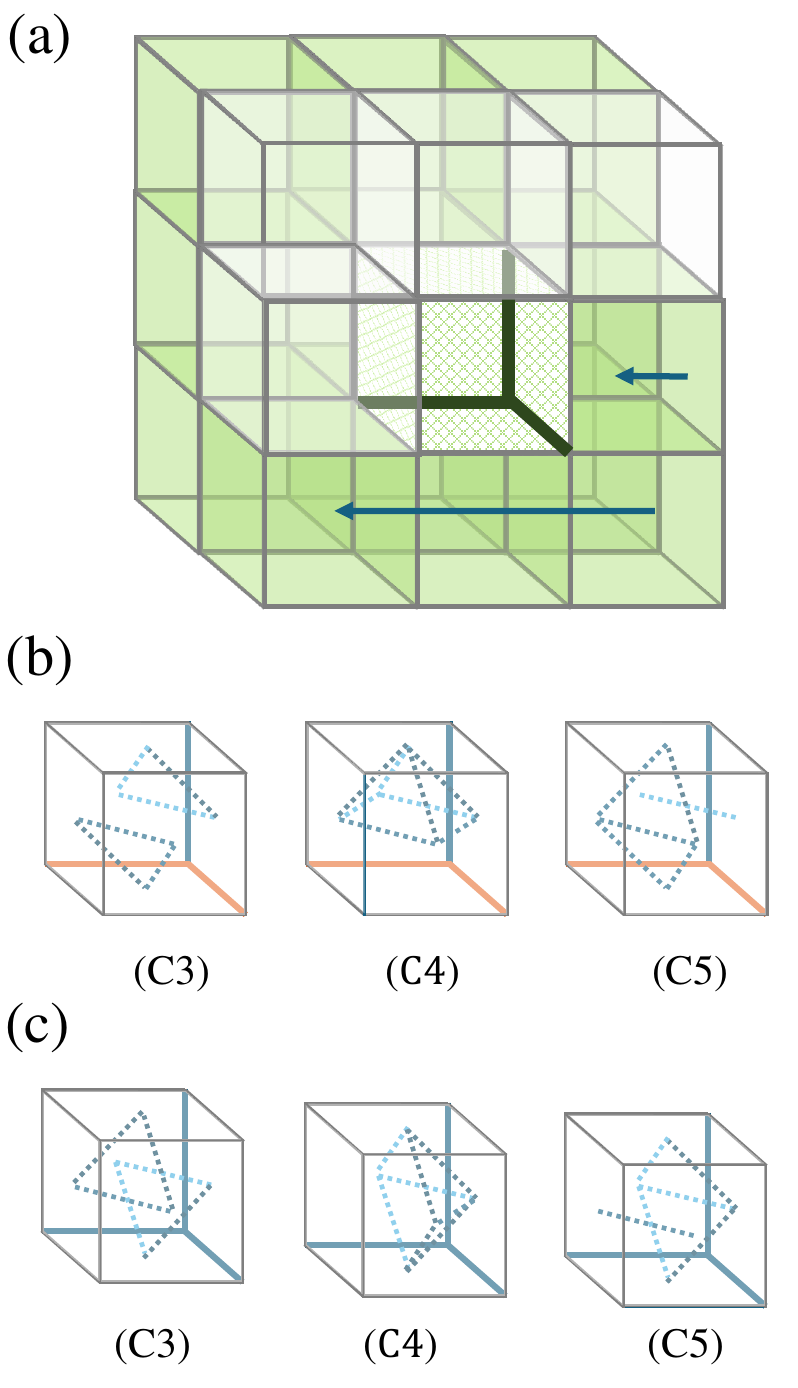}
\caption{\textbf{The scanning method for the cubic lattice.} (a) The orientations of the green blocks have already been assigned. When assigning the orientation of the hatched block, the loops around the three bold edges are completed. Orientations of Blocks~C3, C4, and C5, which contribute one (b) or three (c) hinges (blue edges) to the newly enclosed vertex. The cases in (b) can be rotated to bring the hinge to the needy edge.}
\label{fig:scan 3d}
\end{figure}

Let us examine if the defect set in a 3D metamaterial is subject to any constraints other than having to form closed curves. In the cases of Blocks~C3, C4, and C5, it is not, and this can be seen by a 3D version (Fig.~\ref{fig:scan 3d}a) of the scanning process employed above for the square and honeycomb lattices. Indeed as we construct our metamaterial layer-by-layer, and line-by-line within each layer, each new block completes the neighborhood of one vertex. The parity constraint on the desired defect set means that at this time, the vertex requires one or three hinges to be placed near it. As all three of the said blocks have vertices with one (Fig.~\ref{fig:scan 3d}b), as well as with three (Fig.~\ref{fig:scan 3d}c), incident hinges, a suitable orientation is easy to find.

\subsubsection{Block C6 defect realization}
\label{sec:checkerboard}

\rs{Since in Block~C6 all vertices have exactly one edge with a hinge, this simple scanning method cannot be directly applied to it, and a more delicate procedure is required. \rs{ In Appendix~\ref{sec:c6}} we provide a construction protocol for realizing with Block~C6 any defect configuration which satisfies the parity constraint explained in Sec.~\ref{sec:compatibility} above.}

\subsubsection{Block C2 counting argument for non-realizability}

The following counting argument yields that for large enough systems, most defect sets will not be possible with Blocks~C2, even after restricting attention to configurations with even degree everywhere. The number of metamaterials of size $L \times L \times L$ built from this block is $M=3^{L^3}$, while the number of internal edges is $E=3L(L-1)^2$, and the number of internal vertices is $V=(L-1)^3$. Thus, the number of defect set candidates is $S = \frac{2^E}{2^V} = 2^{(2L+1)(L-1)^2}$, since each internal edge can be either defected or not, and at each internal vertex the parity constraint reduces the set by half. For large $L$, this number of defect set candidates scales as $S \propto 4^{L^3}$, which is larger than the number $M$ of possible structures. Therefore, for large enough lattices, most defect sets are not realizable with Block~C2. By considering the exact expressions for $S$ and $M$ we see that this is certainly the case for $L \ge 8$.

In order to find a defect set which is non-realizable with Block~C2, we wrote a computer program that maps our problem onto an instance of the Boolean satisfiability problem, also known as SAT, to check if a given set is realizable or not~\cite{defect_realization_code}; see Appendix~\ref{sec:code} for more details. To generate candidate defect sets, whose realizability the program would check, we first solved a forward problem of finding the defects in a randomly chosen structure of Blocks~C3. 

For Block~C2, although the counting argument guarantees the existence of non-realizable sets from size $8\times8\times8$, we found that already at size $7\times7\times7$ most candidate sets are non-realizable. On the other hand, we failed at producing small and simple non-realizable examples; Figure~\ref{fig:random geometry knot} below gives an indication of the complexity that seems to be required for non-realizability.

\subsubsection{Block C2 arbitrary knottedness}

We will now show that out of all types of 3D blocks, save for the trivial Block C1, it is possible to build metamaterials in such a way that the defect is a non-self-intersecting closed curve of an arbitrarily chosen knot or link type. This is obvious for Blocks~C3, C4, C5, and C6, where we can pick any polygonal representation of the desired knot or link, built from edges of a large enough lattice, and realize it as the defect with the method outlined in previous subsections. We do not have such complete flexibility with Block~C2. In this case our design will be based on \emph{grid diagrams}, which we now review. 

A grid diagram of order $K$ is given by two collections of $K$ points in the plane, of the form
\begin{multline*}
\mathbf X=\{(1,\xi_1),(2,\xi_2),\ldots,(K,\xi_K)\}\text{ and}\\
\mathbf O=\{(1,\omega_1),(2,\omega_2),\ldots,(K,\omega_K)\},
\end{multline*}
where both $(\xi_1,\ldots,\xi_K)$ and $(\omega_1,\ldots,\omega_K)$ are permutations of the set $\{1,\ldots,K\}$ in such a way that $\xi_i\ne\omega_i$ for all $i=1,\ldots,K$. That is, we have altogether $2K$ nodes of a $(K-1)\times(K-1)$ grid so that every vertical and every horizontal grid line contains exactly one element of $\mathbf X$ and exactly one element of $\mathbf O$.

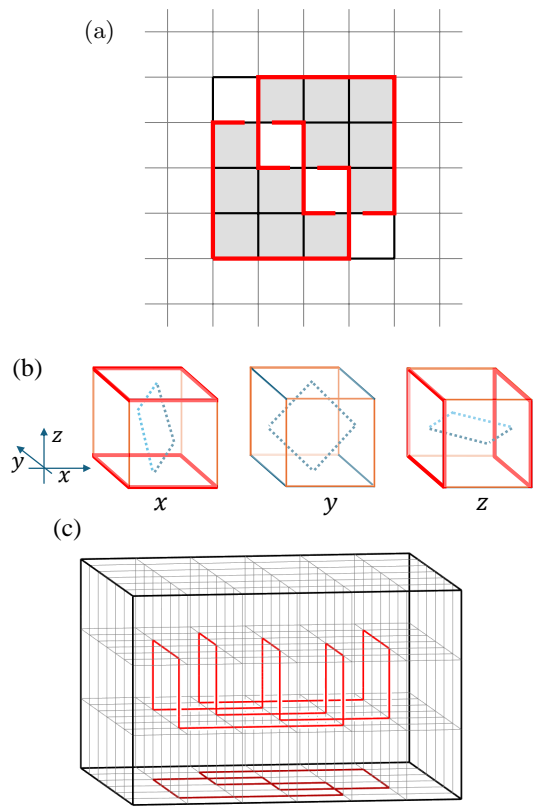
\begin{figure}[t]
\begin{subfigure}[b]{\columnwidth}
\begin{tikzpicture}[scale=.60]
\node at (-0.5, 7) {\small (a)};
%\node at (-0.5, 0) {\small (b)};
\path[fill=lightgray,semitransparent] (2,2)--(5,2)--(5,3)--(4,3)--(4,4)--(3,4)--(3,5)--(2,5)--(2,2);
\path[fill=lightgray,semitransparent] (6,3)--(6,6)--(3,6)--(3,5)--(4,5)--(4,4)--(5,4)--(5,3)--(6,3);
\draw[step=1.0,help lines] (0.5,0.5) grid (7.5,7.5);
\draw[step=1.0,black,thick] (2,2) grid (6,6);
\draw[ultra thick,red] (2,2)--(2,5)--(2.7,5);
\draw[ultra thick,red] (3.3,5)--(4,5)--(4,3)--(4.7,3);
\draw[ultra thick,red] (5.3,3)--(6,3)--(6,6)--(3,6)--((3,4)--(3.7,4);
\draw[ultra thick,red] (4.3,4)--(5,4)--(5,2)--(2,2);
\end{tikzpicture}
\begin{subfigure}[b]{\columnwidth}
\includegraphics[width=.85\columnwidth]{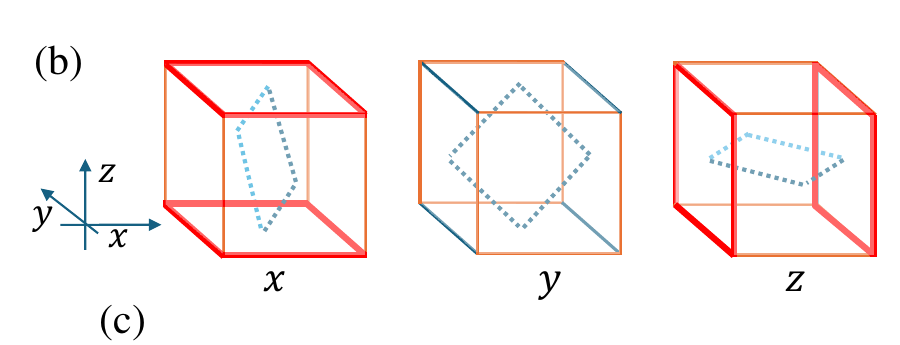}
\end{subfigure}
\end{subfigure}
\begin{subfigure}[b]{\columnwidth}
\includegraphics[width=.63\columnwidth]{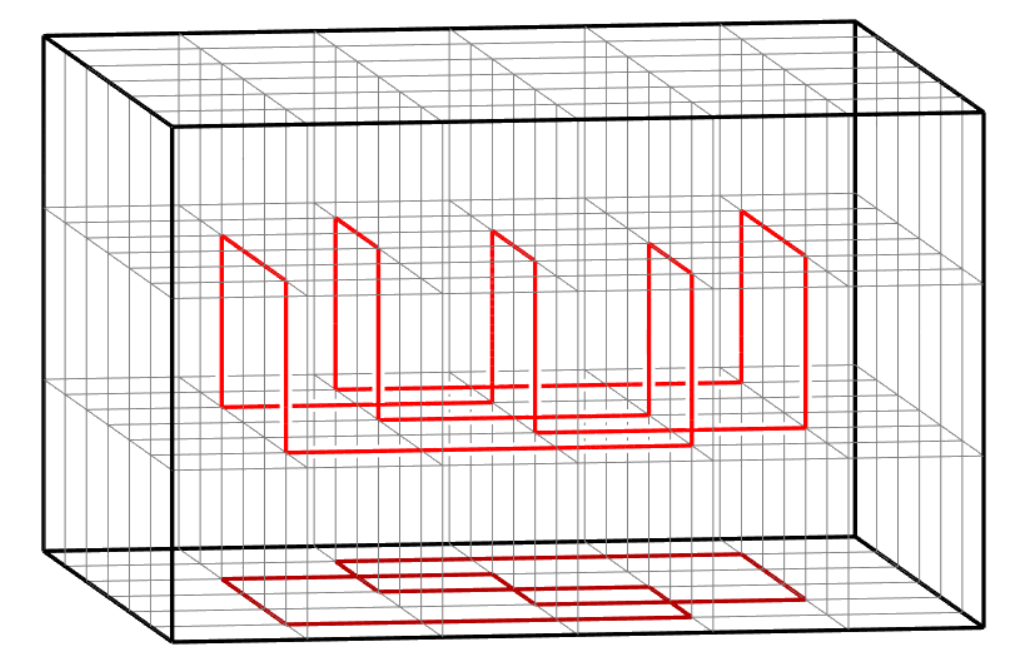}
\end{subfigure}
\caption{\textbf{Realizing arbitrary knottedness with Block~C2.} (a) Grid diagram of the trefoil knot with $K=5$. (b) The three orientations of Block~C2. The red lines mark the defected edges when replacing one unit in an `all $y$' configuration with each of the other orientations. (c) A trefoil defect set in a 3D metamaterial built from Block~C2.}
\label{fig:grid_example}
\end{figure}

Now the grid diagram itself is constructed by connecting all pairs of horizontally or vertically aligned nodes with $2K$ straight line segments. The final and most key point is that whenever two such segments intersect, we consistently make the vertical one cross over the horizontal one, as shown in Fig.~\ref{fig:grid_example}a. It is well known and easy to prove that all knot and link types possess grid diagrams. We also endow our grid diagrams with a checkerboard shading, in which the unbounded region is not shaded.

To build our metamaterial with a knotted defect, we start with a reference state, in which we fill an entire spatial region with identically oriented blocks of type C2. To be concrete, a $(K+1)\times(K+1)\times3$ lattice suffices. The defect set of our reference state is empty, i.e., the initial metamaterial is compatible. This is easy to check in each case, but it can also be argued as follows: among any set of four parallel edges of any block, an even number of them will be struts and therefore another even number will be hinges. Thus when these four edges are brought together by parallel translations, the resulting edge is indeed not defected.

We start creating a defect by reassigning the orientations of certain blocks. Every time we do that, the defectedness of the $12$ incident edges potentially changes. It will actually change exactly when, as a result of reorienting the block, the edge changes from a strut to a hinge, or vice versa. Figure~\ref{fig:grid_example}b shows an exact list of the changes that may occur. We see that for Block~C2, there exists a reorientation so that the newly frustrated edges form the boundaries of two opposite faces of the block. Without loss of generality (by choosing the appropriate reference state) we may assume that the two faces are perpendicular to the $z$-axis. 

Our next step is to apply the above change, along the bottom ($z=-1$) level, to the orientations of the blocks which correspond to the shaded squares of the desired grid diagram. (We place the grid diagram in the middle of our $(K+1)\times(K+1)$ square, as in Fig.~\ref{fig:grid_example}a.) This creates, as the defect set of the metamaterial, a copy of the grid \emph{projection}, with self-intersections. 

In order to build the over- and under-crossings, we make some further orientation changes. Namely, we re-assign the orientations of the same set of blocks we started with (corresponding to the shaded squares), but this time at the middle ($z=0$) level in such a way that each time, the boundaries of two squares, perpendicular to the $x$-axis this time, are added to the defect set, cf.\ Fig.~\ref{fig:grid_example}b. In Fig.~\ref{fig:grid_example}c we show an example of the outcome. In Appendix~\ref{sec:examples} we give specific block assignments which illustrate our method.

\begin{figure}[h!]
\centering
\includegraphics[width=1\columnwidth]{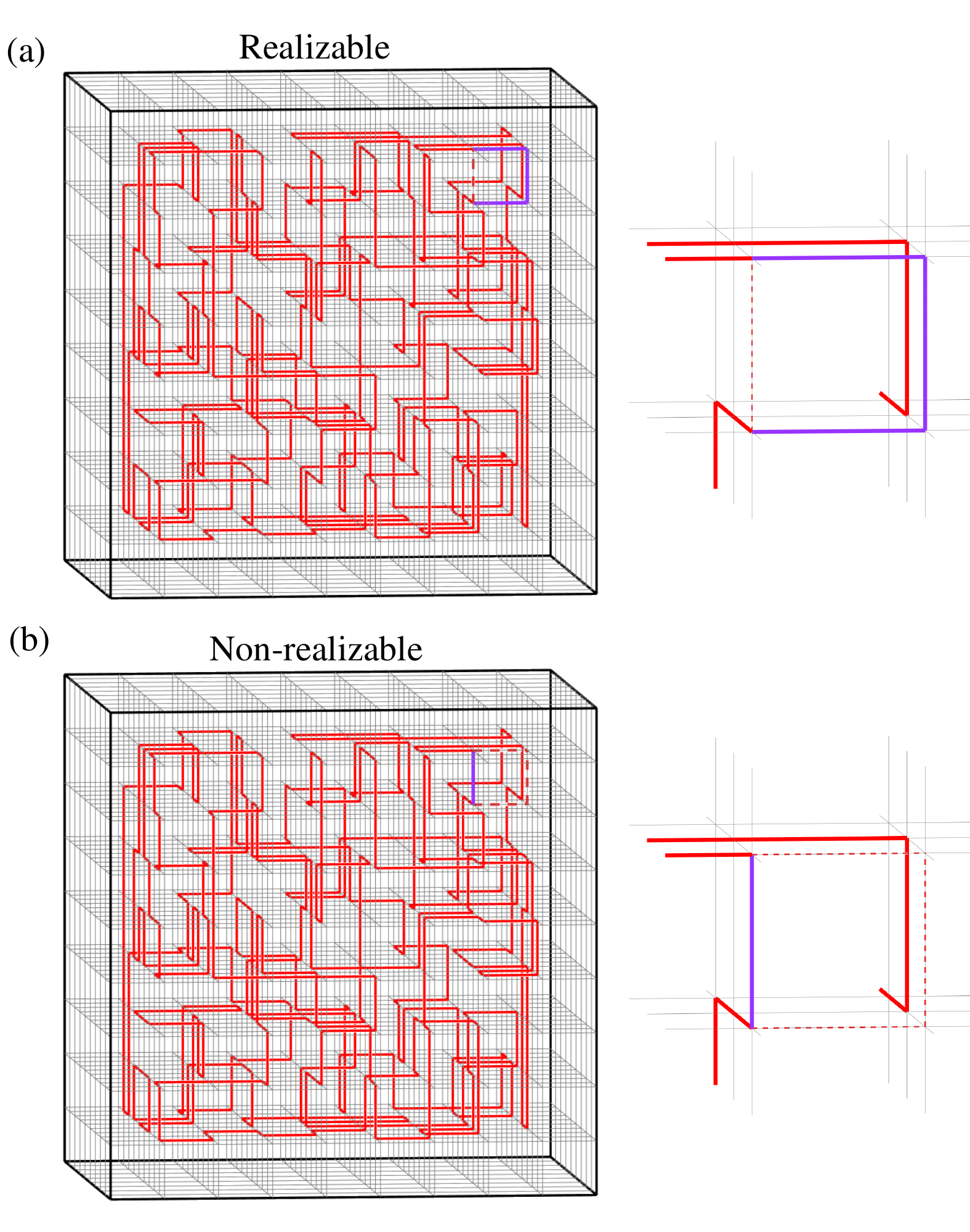}
\caption{\textbf{Trefoil knot with random geometric changes.} (a) realizable and (b)~non-realizable defect set, with Blocks~C2, and a zoomed-in image of the difference between the two (left). The purple lines mark defects in the present set, and the dashed lines are defects in the other set but not in the present one.}
\label{fig:random geometry knot}
\end{figure}

\subsubsection{Block C2 topology versus geometry}

Let us present an example to underscore that it is not the topological complexity (namely knot or link type) of the defect lines that prevents them from being realizable in a combinatorial metamaterial made of Block~C2; rather it is some aspect of their geometric complexity.

We took the basic trefoil knot shown in Fig.~\ref{fig:grid_example}b and randomly changed its geometry. We performed this by iteratively choosing a random internal face within the metamaterial, attempting to invert the defectedness of all four of its edges, and accepting the move only if the topology was not changed, i.e., there were no newly formed intersections or components of defect lines. Given that for completely random defect sets, we found above that with Blocks~C2 and a lattice size of at least $7\times7\times7$ most defect sets are non-realizable, and since in our current candidate sets we assign defects only to edges of internal faces, we looked at a lattice of size $9\times9\times9$ so that the randomized knot could be surrounded by at least one block on each side. After about 700 iterations of the randomization procedure there appeared sets that, although they maintained the trefoil topology, had random enough geometry to render them non-realizable. Figure~\ref{fig:random geometry knot} shows a pair of configurations where a switch from realizable to non-realizable occurs. We expect that after multiple further iterations of this procedure, most defect sets will be non-realizable. 

\section{Discussion}
\label{sec:discussion}

We systematically studied the ability to design metamaterials with a desired spatial distribution of mechanical frustration, in the form of localized defects. Our combinatorial approach allowed us much freedom in positioning defects within the lattice, with arbitrary geometrical positioning of defects for Blocks~S3, S4, H4, H$5_a$, H$5_b$, H6, C3, C4, C5, and C6, and with arbitrary knottedness topology of defect lines with Block~C2. In the accompanying paper~\cite{compatible_paper} we studied the ability to design the deformation texture on the boundary of the metamaterial, and the multiplicity of compatible metamaterials made of the same building blocks for which, in this paper, we studied the ability to position defects. We showed how these properties of compatible metamaterials depend on whether the blocks induce holographic order or not. Here we find that in the presence of holographic order, square Block~S3 and cubic Block~C3 allow any defect set to be realized, while the hexagonal Blocks~H2 and H$3_a$ as well as the cubic Block~C2 do not. Interestingly, blocks that do not induce holographic order, namely S4, H4, H$5_a$, H$5_b$, H6, C4, C5, and C6 all allow any defect set to be realized.

The effect of defects on the mechanical response of metamaterials has been demonstrated\rs{, for instance by stressed regions bypassing topological defects in 2D}~\cite{meeussen2020supertriangles}. It would be interesting to understand how the presence of multiple defects, possibly with a complex spatial distribution, affects the mechanical response. \rs{Realization of metamaterials with non-trivial defect structures, and specifically with complex knotted topologies, is interesting in its own right.} Another direction in this regard would be identifying mechanical consequences stemming from the topology (as in, knottedness) of defect lines within metamaterials, which we are now able to to generate. \rs{Closed loops of defects can cause stress to concentrate in the region enclosed by the loop, or alternatively to avoid this region~\cite{pisanty2021putting}. We are not currently aware of specific mechanical implications for realizing knotted defect lines, however we note that knotted floppy chains have been recently realized in combinatorial metamaterials~\cite{T1T2}, and suggest that it would be interesting to explore, for instance, how linking several defect loops alters stress steering in metamaterials.}

It would also be interesting to extend our approach of generating defect lines with arbitrary knottedness to situations with defect line crossings, where vertices have 4 or 6 defected edges. \rs{Another intriguing possibility would be dynamically repositioning defects --- for example, by using bistable blocks --- in ways that would preserve topological features.} Finally, for the 3D blocks that we considered, defect lines cannot terminate or split in the bulk. It would be interesting to identify metamaterials that could have point defects in 3D, similarly to hedgehog point defects in nematic liquid crystals.

\begin{acknowledgments}

We thank Corentin Coulais, Martin van Hecke, Ben Pisanty, Priyanka, Dor Shohat, Tomer Sigalov, and Ivan Smalyukh for helpful discussions. This research was supported in part by the Israel Science Foundation Grant Nos. 1899/20 and 2902/21. Support was also provided by the Japan Society for the Promotion of Science (JSPS) Grant-in-Aid for Scientific Research C, no.\ 23K03108. C.S.K. was supported by the Clore Scholars Programme. Y.M.Y.F. was supported by Schmidt Science Fellows, in partnership with the Rhodes Trust. 

\end{acknowledgments}

\emph{Author Contributions:} C.S.K., T.K. and Y.S. conceived of and planned the research, performed the theoretical research, and wrote the paper; Y.M.Y.F. and G.C. conceived of and wrote the computer program for 3D defect set realizations, which C.S.K. then used for the computational research.

\appendix

\section{H6 arbitrary defect positioning}
\label{sec:caterpillar}

Block H6 contains two hinges and we will indicate the orientation of the block by connecting the corners of the hinges by an arc. Then a vertex of the lattice is defected if it is the endpoint of an odd number ($1$ or $3$) of such arcs, and it is undefected in the case of an even number ($0$ or $2$) of incident arcs. 

For an arbitrary set $\mathcal{D}$ of vertices, we will design a metamaterial in which exactly the elements of $\mathcal{D}$ are defected. We say that a vertex is needy if the number of its incident arcs has the opposite of the desired parity. At the beginning of the process, before any blocks are filled in, the set of needy vertices is exactly $\mathcal{D}$. As we proceed, we will keep updating the set of needy vertices and we will indicate that a vertex is needy by placing a square mark there. Our goal is that at the end there be no needy vertices at all.

The vertices of the hexagonal lattice form two triangular sublattices, which we will represent by the colors brown and green. Notice that the hinges of each block occur at vertices of the same color, whereby our arcs are also naturally colored brown and green.

Our first basic observation is that in any row of hexagons, if an even number of their vertices of the same color are specified, then there is a unique way to connect them in pairs with sequences of arcs of that color. Let us call this the \emph{caterpillar matching}. See Fig.~\ref{fig:filling} for some examples and note that a caterpillar matching need not specify an arc in every block of the row. Adding a caterpillar matching will change the neediness status of exactly the specified vertices that generated the matching.

A caterpillar matching can also be applied in an infinite row, in fact with any number of vertices to be matched, including odd numbers and infinity. If the row is infinite in both directions, then the number of possibilities is not just one but two.

Suppose now that we have a full row of blocks that are not yet assigned. Along the top boundary of this row we assume not to have the following `bad pattern':
\[\begin{tikzpicture}[scale=.2]
\begin{scope}[rotate=30]
\draw (-1,-1.71)--(0,0)--(4,0)--(5,-1.71);
\draw (4,0)--(5,1.71);
\draw (0,0)--(-2,3.42)--(-4,3.42);
\draw (-2,3.42)--(-1,5.13);
\node at (0,0) {\color{brown}$\blacksquare$};
\node at (4,0) {\color{green}$\blacksquare$};
\node at (-2,3.42) {\color{green}$\blacksquare$};
\end{scope}
\end{tikzpicture}.\]
We shall orient the blocks in this row in such a way that none of the upper vertices remains needy (notice how a single bad pattern would prevent this), and so that along the lower boundary the bad pattern is again avoided.

We sort the hexagons of our row to `green' and `brownish' ones according to whether their top vertex is needy or not. This divides the strip into alternating green and brownish segments, some of which may consist of a single block.

\usetikzlibrary{decorations.pathreplacing}
\begin{figure}[h]
\begin{tikzpicture}[scale=.15]
\draw [decorate,decoration={brace,amplitude=5pt}] (-27,15)--(-7.5,15) node[midway,yshift=12pt]{\small brownish};
\draw [decorate,decoration={brace,amplitude=5pt}] (-6.5,15)--(2.5,15) node[midway,yshift=12pt]{\small green};
\draw [decorate,decoration=
{brace,amplitude=5pt}] (3.5,15)--(13,15) node[midway,yshift=12pt]{\small brownish};
\draw [decorate,decoration={brace,amplitude=5pt}] (14,15)--(27,15) node[midway,yshift=12pt]{\small green};
\begin{scope}[shift={(0,12)},rotate=30]
\path[fill=lightgray,semitransparent,shift={(-24,13.68)}] (2,-3.42)--(0,-3.42)--(-1,-1.71)--(0,0)--(2,0)--(3,-1.71)--(2,-3.42);
\path[fill=lightgray,semitransparent,shift={(-21,11.97)}] (2,-3.42)--(0,-3.42)--(-1,-1.71)--(0,0)--(2,0)--(3,-1.71)--(2,-3.42);
\path[fill=lightgray,semitransparent,shift={(-18,10.26)}] (2,-3.42)--(0,-3.42)--(-1,-1.71)--(0,0)--(2,0)--(3,-1.71)--(2,-3.42);
\path[fill=lightgray,semitransparent,shift={(-15,8.55)}] (2,-3.42)--(0,-3.42)--(-1,-1.71)--(0,0)--(2,0)--(3,-1.71)--(2,-3.42);
\path[fill=lightgray,semitransparent,shift={(-12,6.84)}] (2,-3.42)--(0,-3.42)--(-1,-1.71)--(0,0)--(2,0)--(3,-1.71)--(2,-3.42);
\path[fill=lightgray,semitransparent,shift={(-9,5.13)}] (2,-3.42)--(0,-3.42)--(-1,-1.71)--(0,0)--(2,0)--(3,-1.71)--(2,-3.42);
\path[fill=lightgray,semitransparent,shift={(3,-1.71)}] (2,-3.42)--(0,-3.42)--(-1,-1.71)--(0,0)--(2,0)--(3,-1.71)--(2,-3.42);
\path[fill=lightgray,semitransparent,shift={(6,-3.42)}] (2,-3.42)--(0,-3.42)--(-1,-1.71)--(0,0)--(2,0)--(3,-1.71)--(2,-3.42);
\path[fill=lightgray,semitransparent,shift={(9,-5.13)}] (2,-3.42)--(0,-3.42)--(-1,-1.71)--(0,0)--(2,0)--(3,-1.71)--(2,-3.42);
\draw[shift={(-24,13.68)}] (2,-3.42)--(0,-3.42)--(-1,-1.71)--(0,0)--(2,0)--(3,-1.71);
\node at (-24,10.26) {\color{brown}\tiny$\blacksquare$};
\draw[shift={(-21,11.97)}] (2,-3.42)--(0,-3.42)--(-1,-1.71)--(0,0)--(2,0)--(3,-1.71);
\draw[shift={(-18,10.26)}] (2,-3.42)--(0,-3.42)--(-1,-1.71)--(0,0)--(2,0)--(3,-1.71);
\node at (-18,6.84) {\color{brown}\tiny$\blacksquare$};
\node at (-19,8.55) {\color{green}\tiny$\blacksquare$};
\draw[shift={(-15,8.55)}] (2,-3.42)--(0,-3.42)--(-1,-1.71)--(0,0)--(2,0)--(3,-1.71);
\node at (-15,8.55) {\color{brown}\tiny$\blacksquare$};
\node at (-16,6.84) {\color{green}\tiny$\blacksquare$};
\draw[shift={(-12,6.84)}] (2,-3.42)--(0,-3.42)--(-1,-1.71)--(0,0)--(2,0)--(3,-1.71);
\node at (-12,6.84) {\color{brown}\tiny$\blacksquare$};
\draw[shift={(-9,5.13)}] (2,-3.42)--(0,-3.42)--(-1,-1.71)--(0,0)--(2,0)--(3,-1.71);
\node at (-9,1.71) {\color{brown}\tiny$\blacksquare$};
\node at (-10,3.42) {\color{green}\tiny$\blacksquare$};
\draw[shift={(-6,3.42)}] (2,-3.42)--(0,-3.42)--(-1,-1.71)--(0,0)--(2,0)--(3,-1.71);
\node at (-6,0) {\color{brown}\tiny$\blacksquare$};
\node at (-4,3.42) {\color{green}\tiny$\blacksquare$};
\draw[shift={(-3,1.71)}] (2,-3.42)--(0,-3.42)--(-1,-1.71)--(0,0)--(2,0)--(3,-1.71);
\node at (-1,1.71) {\color{green}\tiny$\blacksquare$};
\draw (2,-3.42)--(0,-3.42)--(-1,-1.71)--(0,0)--(2,0)--(3,-1.71);
\node at (0,-3.42) {\color{brown}\tiny$\blacksquare$};
\node at (2,0) {\color{green}\tiny$\blacksquare$};
\draw[shift={(3,-1.71)}] (2,-3.42)--(0,-3.42)--(-1,-1.71)--(0,0)--(2,0)--(3,-1.71);
\node at (3,-1.71) {\color{brown}\tiny$\blacksquare$};
\node at (3,-5.13) {\color{brown}\tiny$\blacksquare$};
\node at (2,-3.42) {\color{green}\tiny$\blacksquare$};
\draw[shift={(6,-3.42)}] (2,-3.42)--(0,-3.42)--(-1,-1.71)--(0,0)--(2,0)--(3,-1.71);
\node at (5,-5.13) {\color{green}\tiny$\blacksquare$};
\draw[shift={(9,-5.13)}] (2,-3.42)--(0,-3.42)--(-1,-1.71)--(0,0)--(2,0)--(3,-1.71);
\draw[shift={(12,-6.84)}] (2,-3.42)--(0,-3.42)--(-1,-1.71)--(0,0)--(2,0)--(3,-1.71);
\node at (12,-6.84) {\color{brown}\tiny$\blacksquare$};
\node at (12,-10.26) {\color{brown}\tiny$\blacksquare$};
\node at (11,-8.55) {\color{green}\tiny$\blacksquare$};
\node at (14,-6.84) {\color{green}\tiny$\blacksquare$};
\draw[shift={(15,-8.55)}] (2,-3.42)--(0,-3.42)--(-1,-1.71)--(0,0)--(2,0)--(3,-1.71);
\node at (17,-8.55) {\color{green}\tiny$\blacksquare$};
\draw[shift={(18,-10.26)}] (2,-3.42)--(0,-3.42)--(-1,-1.71)--(0,0)--(2,0)--(3,-1.71);
\node at (17,-11.97) {\color{green}\tiny$\blacksquare$};
\node at (20,-10.26) {\color{green}\tiny$\blacksquare$};
\node at (18,-13.68) {\color{brown}\tiny$\blacksquare$};
\draw[shift={(21,-11.97)}] (2,-3.42)--(0,-3.42)--(-1,-1.71)--(0,0)--(2,0)--(3,-1.71);
\node at (23,-11.97) {\color{green}\tiny$\blacksquare$};
\draw (23,-15.39)--(24,-13.68);
\end{scope}
\draw[->,thick] (-20,8)--(-20,2) node[midway,right] {\small caterpillar matching in the brownish segments}; 
\begin{scope}[rotate=30]
\path[fill=lightgray,semitransparent,shift={(-24,13.68)}] (2,-3.42)--(0,-3.42)--(-1,-1.71)--(0,0)--(2,0)--(3,-1.71)--(2,-3.42);
\draw[shift={(-24,13.68)}] (2,-3.42)--(0,-3.42)--(-1,-1.71)--(0,0)--(2,0)--(3,-1.71);
\draw[thick,brown,shift={(-24,13.68)}] (3,-1.71) arc (90:150:3.42);
\path[fill=lightgray,semitransparent,shift={(-21,11.97)}] (2,-3.42)--(0,-3.42)--(-1,-1.71)--(0,0)--(2,0)--(3,-1.71)--(2,-3.42);
\draw[shift={(-21,11.97)}] (2,-3.42)--(0,-3.42)--(-1,-1.71)--(0,0)--(2,0)--(3,-1.71);
\draw[thick,brown,shift={(-21,11.97)}] (0,0) arc (210:270:3.42);
\path[fill=lightgray,semitransparent,shift={(-18,10.26)}] (2,-3.42)--(0,-3.42)--(-1,-1.71)--(0,0)--(2,0)--(3,-1.71)--(2,-3.42);
\draw[shift={(-18,10.26)}] (2,-3.42)--(0,-3.42)--(-1,-1.71)--(0,0)--(2,0)--(3,-1.71);
\draw[thick,brown,shift={(-18,10.26)}] (0,0) arc (30:-30:3.42);
\node at (-19,8.55) {\color{green}\tiny$\blacksquare$};
\path[fill=lightgray,semitransparent,shift={(-15,8.55)}] (2,-3.42)--(0,-3.42)--(-1,-1.71)--(0,0)--(2,0)--(3,-1.71)--(2,-3.42);
\draw[shift={(-15,8.55)}] (2,-3.42)--(0,-3.42)--(-1,-1.71)--(0,0)--(2,0)--(3,-1.71);
\draw[thick,brown,shift={(-15,8.55)}] (0,0) arc (210:270:3.42);
\node at (-16,6.84) {\color{green}\tiny$\blacksquare$};
\path[fill=lightgray,semitransparent,shift={(-12,6.84)}] (2,-3.42)--(0,-3.42)--(-1,-1.71)--(0,0)--(2,0)--(3,-1.71)--(2,-3.42);
\draw[shift={(-12,6.84)}] (2,-3.42)--(0,-3.42)--(-1,-1.71)--(0,0)--(2,0)--(3,-1.71);
\path[fill=lightgray,semitransparent,shift={(-9,5.13)}] (2,-3.42)--(0,-3.42)--(-1,-1.71)--(0,0)--(2,0)--(3,-1.71)--(2,-3.42);
\draw[shift={(-9,5.13)}] (2,-3.42)--(0,-3.42)--(-1,-1.71)--(0,0)--(2,0)--(3,-1.71);
\node at (-9,1.71) {\color{brown}\tiny$\blacksquare$};
\node at (-10,3.42) {\color{green}\tiny$\blacksquare$};
\draw[shift={(-6,3.42)}] (2,-3.42)--(0,-3.42)--(-1,-1.71)--(0,0)--(2,0)--(3,-1.71);
\node at (-6,0) {\color{brown}\tiny$\blacksquare$};
\node at (-4,3.42) {\color{green}\tiny$\blacksquare$};
\draw[shift={(-3,1.71)}] (2,-3.42)--(0,-3.42)--(-1,-1.71)--(0,0)--(2,0)--(3,-1.71);
\node at (-1,1.71) {\color{green}\tiny$\blacksquare$};
\draw (2,-3.42)--(0,-3.42)--(-1,-1.71)--(0,0)--(2,0)--(3,-1.71);
\node at (0,-3.42) {\color{brown}\tiny$\blacksquare$};
\node at (2,0) {\color{green}\tiny$\blacksquare$};
\path[fill=lightgray,semitransparent,shift={(3,-1.71)}] (2,-3.42)--(0,-3.42)--(-1,-1.71)--(0,0)--(2,0)--(3,-1.71)--(2,-3.42);
\draw[shift={(3,-1.71)}] (2,-3.42)--(0,-3.42)--(-1,-1.71)--(0,0)--(2,0)--(3,-1.71);
\draw[thick,brown,shift={(3,-1.71))}] (0,0) arc (30:-30:3.42);
\node at (2,-3.42) {\color{green}\tiny$\blacksquare$};
\path[fill=lightgray,semitransparent,shift={(6,-3.42)}] (2,-3.42)--(0,-3.42)--(-1,-1.71)--(0,0)--(2,0)--(3,-1.71)--(2,-3.42);
\draw[shift={(6,-3.42)}] (2,-3.42)--(0,-3.42)--(-1,-1.71)--(0,0)--(2,0)--(3,-1.71);
\node at (5,-5.13) {\color{green}\tiny$\blacksquare$};
\path[fill=lightgray,semitransparent,shift={(9,-5.13)}] (2,-3.42)--(0,-3.42)--(-1,-1.71)--(0,0)--(2,0)--(3,-1.71)--(2,-3.42);
\draw[shift={(9,-5.13)}] (2,-3.42)--(0,-3.42)--(-1,-1.71)--(0,0)--(2,0)--(3,-1.71);
\draw[thick,brown,shift={(9,-5.13)}] (3,-1.71) arc (90:150:3.42);
\node at (9,-8.55) {\color{brown}\tiny$\blacksquare$};
\draw[shift={(12,-6.84)}] (2,-3.42)--(0,-3.42)--(-1,-1.71)--(0,0)--(2,0)--(3,-1.71);
\node at (12,-10.26) {\color{brown}\tiny$\blacksquare$};
\node at (11,-8.55) {\color{green}\tiny$\blacksquare$};
\node at (14,-6.84) {\color{green}\tiny$\blacksquare$};
\draw[shift={(15,-8.55)}] (2,-3.42)--(0,-3.42)--(-1,-1.71)--(0,0)--(2,0)--(3,-1.71);
\node at (17,-8.55) {\color{green}\tiny$\blacksquare$};
\draw[shift={(18,-10.26)}] (2,-3.42)--(0,-3.42)--(-1,-1.71)--(0,0)--(2,0)--(3,-1.71);
\node at (17,-11.97) {\color{green}\tiny$\blacksquare$};
\node at (20,-10.26) {\color{green}\tiny$\blacksquare$};
\node at (18,-13.68) {\color{brown}\tiny$\blacksquare$};
\draw[shift={(21,-11.97)}] (2,-3.42)--(0,-3.42)--(-1,-1.71)--(0,0)--(2,0)--(3,-1.71);
\node at (23,-11.97) {\color{green}\tiny$\blacksquare$};
\draw (23,-15.39)--(24,-13.68);
\end{scope}
\draw[->,thick] (-20,-4)--(-20,-10) node[midway,right] {\small filling the gaps in the brownish segments}; 
\begin{scope}[shift={(0,-12)},rotate=30]
\path[fill=lightgray,semitransparent,shift={(-24,13.68)}] (2,-3.42)--(0,-3.42)--(-1,-1.71)--(0,0)--(2,0)--(3,-1.71)--(2,-3.42);
\draw[shift={(-24,13.68)}] (2,-3.42)--(0,-3.42)--(-1,-1.71)--(0,0)--(2,0)--(3,-1.71);
\draw[thick,brown,shift={(-24,13.68)}] (3,-1.71) arc (90:150:3.42);
\path[fill=lightgray,semitransparent,shift={(-21,11.97)}] (2,-3.42)--(0,-3.42)--(-1,-1.71)--(0,0)--(2,0)--(3,-1.71)--(2,-3.42);
\draw[shift={(-21,11.97)}] (2,-3.42)--(0,-3.42)--(-1,-1.71)--(0,0)--(2,0)--(3,-1.71);
\draw[thick,brown,shift={(-21,11.97)}] (0,0) arc (210:270:3.42);
\path[fill=lightgray,semitransparent,shift={(-18,10.26)}] (2,-3.42)--(0,-3.42)--(-1,-1.71)--(0,0)--(2,0)--(3,-1.71)--(2,-3.42);
\draw[shift={(-18,10.26)}] (2,-3.42)--(0,-3.42)--(-1,-1.71)--(0,0)--(2,0)--(3,-1.71);
\draw[thick,brown,shift={(-18,10.26)}] (0,0) arc (30:-30:3.42);
\node at (-19,8.55) {\color{green}\tiny$\blacksquare$};
\path[fill=lightgray,semitransparent,shift={(-15,8.55)}] (2,-3.42)--(0,-3.42)--(-1,-1.71)--(0,0)--(2,0)--(3,-1.71)--(2,-3.42);
\draw[shift={(-15,8.55)}] (2,-3.42)--(0,-3.42)--(-1,-1.71)--(0,0)--(2,0)--(3,-1.71);
\draw[thick,brown,shift={(-15,8.55)}] (0,0) arc (210:270:3.42);
\node at (-16,6.84) {\color{green}\tiny$\blacksquare$};
\path[fill=lightgray,semitransparent,shift={(-12,6.84)}] (2,-3.42)--(0,-3.42)--(-1,-1.71)--(0,0)--(2,0)--(3,-1.71)--(2,-3.42);
\draw[shift={(-12,6.84)}] (2,-3.42)--(0,-3.42)--(-1,-1.71)--(0,0)--(2,0)--(3,-1.71);
\draw[thick,green,shift={(-12,6.84)}] (-1,-1.71) arc (90:30:3.42);
\node at (-13,5.13) {\color{green}\tiny$\blacksquare$};
\path[fill=lightgray,semitransparent,shift={(-9,5.13)}] (2,-3.42)--(0,-3.42)--(-1,-1.71)--(0,0)--(2,0)--(3,-1.71)--(2,-3.42);
\draw[shift={(-9,5.13)}] (2,-3.42)--(0,-3.42)--(-1,-1.71)--(0,0)--(2,0)--(3,-1.71);
\draw[thick,green,shift={(-9,5.13)}] (-1,-1.71) arc (90:30:3.42);
\node at (-9,1.71) {\color{brown}\tiny$\blacksquare$};
\node at (-10,3.42) {\color{green}\tiny$\blacksquare$};
\draw[shift={(-6,3.42)}] (2,-3.42)--(0,-3.42)--(-1,-1.71)--(0,0)--(2,0)--(3,-1.71);
\node at (-6,0) {\color{brown}\tiny$\blacksquare$};
\node at (-7,1.71) {\color{green}\tiny$\blacksquare$};
\node at (-4,3.42) {\color{green}\tiny$\blacksquare$};
\draw[shift={(-3,1.71)}] (2,-3.42)--(0,-3.42)--(-1,-1.71)--(0,0)--(2,0)--(3,-1.71);
\node at (-1,1.71) {\color{green}\tiny$\blacksquare$};
\draw (2,-3.42)--(0,-3.42)--(-1,-1.71)--(0,0)--(2,0)--(3,-1.71);
\node at (0,-3.42) {\color{brown}\tiny$\blacksquare$};
\node at (2,0) {\color{green}\tiny$\blacksquare$};
\path[fill=lightgray,semitransparent,shift={(3,-1.71)}] (2,-3.42)--(0,-3.42)--(-1,-1.71)--(0,0)--(2,0)--(3,-1.71)--(2,-3.42);
\draw[shift={(3,-1.71)}] (2,-3.42)--(0,-3.42)--(-1,-1.71)--(0,0)--(2,0)--(3,-1.71);
\draw[thick,brown,shift={(3,-1.71))}] (0,0) arc (30:-30:3.42);
\node at (2,-3.42) {\color{green}\tiny$\blacksquare$};
\path[fill=lightgray,semitransparent,shift={(6,-3.42)}] (2,-3.42)--(0,-3.42)--(-1,-1.71)--(0,0)--(2,0)--(3,-1.71)--(2,-3.42);
\draw[shift={(6,-3.42)}] (2,-3.42)--(0,-3.42)--(-1,-1.71)--(0,0)--(2,0)--(3,-1.71);
\draw[thick,green,shift={(6,-3.42)}] (-1,-1.71) arc (90:30:3.42);
\path[fill=lightgray,semitransparent,shift={(9,-5.13)}] (2,-3.42)--(0,-3.42)--(-1,-1.71)--(0,0)--(2,0)--(3,-1.71)--(2,-3.42);
\draw[shift={(9,-5.13)}] (2,-3.42)--(0,-3.42)--(-1,-1.71)--(0,0)--(2,0)--(3,-1.71);
\draw[thick,brown,shift={(9,-5.13)}] (3,-1.71) arc (90:150:3.42);
\node at (8,-6.84) {\color{green}\tiny$\blacksquare$};
\node at (9,-8.55) {\color{brown}\tiny$\blacksquare$};
\draw[shift={(12,-6.84)}] (2,-3.42)--(0,-3.42)--(-1,-1.71)--(0,0)--(2,0)--(3,-1.71);
\node at (12,-10.26) {\color{brown}\tiny$\blacksquare$};
\node at (11,-8.55) {\color{green}\tiny$\blacksquare$};
\node at (14,-6.84) {\color{green}\tiny$\blacksquare$};
\draw[shift={(15,-8.55)}] (2,-3.42)--(0,-3.42)--(-1,-1.71)--(0,0)--(2,0)--(3,-1.71);
\node at (17,-8.55) {\color{green}\tiny$\blacksquare$};
\draw[shift={(18,-10.26)}] (2,-3.42)--(0,-3.42)--(-1,-1.71)--(0,0)--(2,0)--(3,-1.71);
\node at (17,-11.97) {\color{green}\tiny$\blacksquare$};
\node at (20,-10.26) {\color{green}\tiny$\blacksquare$};
\node at (18,-13.68) {\color{brown}\tiny$\blacksquare$};
\draw[shift={(21,-11.97)}] (2,-3.42)--(0,-3.42)--(-1,-1.71)--(0,0)--(2,0)--(3,-1.71);
\node at (23,-11.97) {\color{green}\tiny$\blacksquare$};
\draw (23,-15.39)--(24,-13.68);
\end{scope}
\draw[->,thick] (-20,-16)--(-20,-22) node[midway,right] {\small caterpillar matching in the green segments}; 
\begin{scope}[shift={(0,-24)},rotate=30]
\path[fill=lightgray,semitransparent,shift={(-24,13.68)}] (2,-3.42)--(0,-3.42)--(-1,-1.71)--(0,0)--(2,0)--(3,-1.71)--(2,-3.42);
\path[fill=lightgray,semitransparent,shift={(-21,11.97)}] (2,-3.42)--(0,-3.42)--(-1,-1.71)--(0,0)--(2,0)--(3,-1.71)--(2,-3.42);
\path[fill=lightgray,semitransparent,shift={(-18,10.26)}] (2,-3.42)--(0,-3.42)--(-1,-1.71)--(0,0)--(2,0)--(3,-1.71)--(2,-3.42);
\path[fill=lightgray,semitransparent,shift={(-15,8.55)}] (2,-3.42)--(0,-3.42)--(-1,-1.71)--(0,0)--(2,0)--(3,-1.71)--(2,-3.42);
\path[fill=lightgray,semitransparent,shift={(-12,6.84)}] (2,-3.42)--(0,-3.42)--(-1,-1.71)--(0,0)--(2,0)--(3,-1.71)--(2,-3.42);
\path[fill=lightgray,semitransparent,shift={(-9,5.13)}] (2,-3.42)--(0,-3.42)--(-1,-1.71)--(0,0)--(2,0)--(3,-1.71)--(2,-3.42);
\path[fill=lightgray,semitransparent,shift={(3,-1.71)}] (2,-3.42)--(0,-3.42)--(-1,-1.71)--(0,0)--(2,0)--(3,-1.71)--(2,-3.42);
\path[fill=lightgray,semitransparent,shift={(6,-3.42)}] (2,-3.42)--(0,-3.42)--(-1,-1.71)--(0,0)--(2,0)--(3,-1.71)--(2,-3.42);
\path[fill=lightgray,semitransparent,shift={(9,-5.13)}] (2,-3.42)--(0,-3.42)--(-1,-1.71)--(0,0)--(2,0)--(3,-1.71)--(2,-3.42);
\draw[shift={(-24,13.68)}] (2,-3.42)--(0,-3.42)--(-1,-1.71)--(0,0)--(2,0)--(3,-1.71);
\draw[thick,brown,shift={(-24,13.68)}] (3,-1.71) arc (90:150:3.42);
\draw[shift={(-21,11.97)}] (2,-3.42)--(0,-3.42)--(-1,-1.71)--(0,0)--(2,0)--(3,-1.71);
\draw[thick,brown,shift={(-21,11.97)}] (0,0) arc (210:270:3.42);
\draw[shift={(-18,10.26)}] (2,-3.42)--(0,-3.42)--(-1,-1.71)--(0,0)--(2,0)--(3,-1.71);
\draw[thick,brown,shift={(-18,10.26)}] (0,0) arc (30:-30:3.42);
\node at (-19,8.55) {\color{green}\tiny$\blacksquare$};
\draw[shift={(-15,8.55)}] (2,-3.42)--(0,-3.42)--(-1,-1.71)--(0,0)--(2,0)--(3,-1.71);
\draw[thick,brown,shift={(-15,8.55)}] (0,0) arc (210:270:3.42);
\node at (-16,6.84) {\color{green}\tiny$\blacksquare$};
\draw[shift={(-12,6.84)}] (2,-3.42)--(0,-3.42)--(-1,-1.71)--(0,0)--(2,0)--(3,-1.71);
\draw[thick,green,shift={(-12,6.84)}] (-1,-1.71) arc (90:30:3.42);
\node at (-13,5.13) {\color{green}\tiny$\blacksquare$};
\draw[shift={(-9,5.13)}] (2,-3.42)--(0,-3.42)--(-1,-1.71)--(0,0)--(2,0)--(3,-1.71);
\draw[thick,green,shift={(-9,5.13)}] (-1,-1.71) arc (90:30:3.42);
\node at (-9,1.71) {\color{brown}\tiny$\blacksquare$};
\node at (-10,3.42) {\color{green}\tiny$\blacksquare$};
\draw[shift={(-6,3.42)}] (2,-3.42)--(0,-3.42)--(-1,-1.71)--(0,0)--(2,0)--(3,-1.71);
\draw[thick,green,shift={(-6,3.42)}] (2,0) arc (-30:-90:3.42);
\node at (-6,0) {\color{brown}\tiny$\blacksquare$};
\draw[shift={(-3,1.71)}] (2,-3.42)--(0,-3.42)--(-1,-1.71)--(0,0)--(2,0)--(3,-1.71);
\draw[thick,green,shift={(-3,1.71)}] (2,0) arc (150:210:3.42);
\draw (2,-3.42)--(0,-3.42)--(-1,-1.71)--(0,0)--(2,0)--(3,-1.71);
\draw[thick,green] (2,0) arc (-30:-90:3.42);
\node at (0,-3.42) {\color{brown}\tiny$\blacksquare$};
\draw[shift={(3,-1.71)}] (2,-3.42)--(0,-3.42)--(-1,-1.71)--(0,0)--(2,0)--(3,-1.71);
\draw[thick,brown,shift={(3,-1.71))}] (0,0) arc (30:-30:3.42);
\node at (2,-3.42) {\color{green}\tiny$\blacksquare$};
\draw[shift={(6,-3.42)}] (2,-3.42)--(0,-3.42)--(-1,-1.71)--(0,0)--(2,0)--(3,-1.71);
\draw[thick,green,shift={(6,-3.42)}] (-1,-1.71) arc (90:30:3.42);
\draw[shift={(9,-5.13)}] (2,-3.42)--(0,-3.42)--(-1,-1.71)--(0,0)--(2,0)--(3,-1.71);
\draw[thick,brown,shift={(9,-5.13)}] (3,-1.71) arc (90:150:3.42);
\node at (9,-8.55) {\color{brown}\tiny$\blacksquare$};
\node at (8,-6.84) {\color{green}\tiny$\blacksquare$};
\draw[shift={(12,-6.84)}] (2,-3.42)--(0,-3.42)--(-1,-1.71)--(0,0)--(2,0)--(3,-1.71);
\draw[thick,green,shift={(12,-6.84)}] (2,0) arc (-30:-90:3.42);
\node at (12,-10.26) {\color{brown}\tiny$\blacksquare$};
\draw[shift={(15,-8.55)}] (2,-3.42)--(0,-3.42)--(-1,-1.71)--(0,0)--(2,0)--(3,-1.71);
\draw[thick,green,shift={(15,-8.55)}] (2,0) arc (150:210:3.42);
\draw[shift={(18,-10.26)}] (2,-3.42)--(0,-3.42)--(-1,-1.71)--(0,0)--(2,0)--(3,-1.71);
\draw[thick,green,shift={(18,-10.26)}] (2,0) arc (150:210:3.42);
\node at (18,-13.68) {\color{brown}\tiny$\blacksquare$};
\draw[shift={(21,-11.97)}] (2,-3.42)--(0,-3.42)--(-1,-1.71)--(0,0)--(2,0)--(3,-1.71);
\draw[thick,green,shift={(21,-11.97)}] (2,0) arc (-30:-90:3.42);
\draw (23,-15.39)--(24,-13.68);
\end{scope}
\end{tikzpicture}
\caption{\textbf{Constructing a metamaterial layer, with prescribed defects, out of Block~H6.}}
\label{fig:filling}
\end{figure}
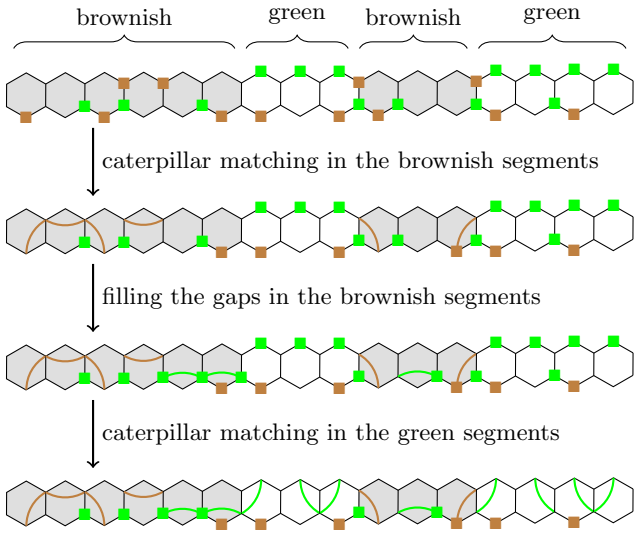

As illustrated in Fig.~\ref{fig:filling}, first we consider the brownish segments and perform a caterpillar matching of their needy brown vertices. If there is an odd number of such, then we add or remove the lower right brown vertex to or from the set. After this, we fill in any remaining hexagons of the brownish segment by green diagonals that connect the two lower green vertices. Notice that there are no needy green vertices along the upper boundary of a brownish segment, neither before nor after constructing its blocks.

Having thus updated the set of needy vertices, we now turn to the green segments and note that they do not have needy brown vertices along their upper boundary. Indeed, the leftmost and rightmost ones have just been taken care of and any others do not exist because they would be part of a bad pattern. Therefore it makes sense to construct a caterpillar matching of all the needy green vertices along the boundary of the green segment. Again, if the number is odd, then we modify the set by adding or removing the lower right green vertex. This step in the construction uses all blocks of the green segment and therefore the row is complete.

It is clear that we have not left any needy vertices along the upper boundary. Regarding possible bad patterns along the lower boundary, notice that a needy brown vertex there can be of two types. If it occurs at the lower right of a brownish segment, then the green vertex to its right belongs to a green segment and therefore it is not needy. If the needy brown vertex is in a green segment, then the green vertex immediately to its left cannot be needy. Therefore as intended, bad patterns do not exist along the lower boundary.

The above construction of a row is of course tailor-made for iteration through consecutive rows in the lattice; all that is left is to get such a process started. For that, pick any horizontal row and perform a caterpillar matching of the brown elements of $\mathcal{D}$ along its boundary, followed by filling the remaining hexagons by arbitrary green arcs. After this, all needy vertices along the boundary will be green and therefore there will not be any bad patterns along the lower boundary. This allows us to use our iterative process to fill the entire half-plane below our starting strip. In fact by symmetry, the same can be said about the upper half-plane too; note that here the bad pattern to be avoided involves a green vertex flanked by two brown ones, i.e., the roles of the two colors flip.

This completes our analysis of Block H6; by our general duality argument, with this we have also established that arbitrary defect sets can be realized with Blocks H5$_a$ as well.

\section{Defects realization with Block~C6}
\label{sec:c6}

\rs{Since in Block~C6 all vertices have exactly one edge with a hinge, a simple scanning method cannot be directly applied to it, and a more delicate procedure is required. Here we provide a construction protocol for realizing with Block~C6 any defect configuration, provided that it  satisfies the parity constraint explained in Sec.~\ref{sec:compatibility} in the main text. Namely, the number of defected edges meeting at any vertex inside the lattice must be even. A useful practical consequence of this is that if we design the metamaterial while making sure that the defectedness of five of the six local edges is as in the desired defect configuration, then the parity constraint will assure that the sixth edge fits the prescription, too.}

\rs{We begin with the top $z$ layer of the metamaterial, and for it we will take care of the defectedness only of the edges in the $z$ direction within it. To achieve this, we will use the fact that the 3D Block~C6 is comprised of three 2D square blocks in the three orthogonal directions, one of which is Block~S2 and the other two are~S4. (A similar decomposition of the other cubic block types is given and utilized in the accompanying paper on compatible metamaterials made of the block types studied here~\cite{compatible_paper}.) Therefore, we can orient the blocks in the top $z$ layer such that their behavior in the $xy$ plane is that of square Block~S4, for which we showed above that we can realize any defect configuration. The defected vertices in the 2D $xy$ plane now extend to defected edges in the $z$ direction going through this layer.}

\rs{Next, we iteratively assign block orientations in $z$ layers going from this top layer all the way down to the bottom $z$ layer. When treating each such layer, we will make sure to satisfy the desired defectedness of the edges in the $x$ and $y$ directions that are in the $xy$ plane between this layer and the previous one. In this way, at each vertex between this newly assigned $z$ layer and the layer above it, the edge going in the positive $z$ direction already has the desired defectedness due to the resolution of the previous layer, and the remaining edge going in the negative $z$ direction will have the desired defectedness due to parity considerations.}

\begin{figure}[t]
\centering
\includegraphics[width=.8\columnwidth]{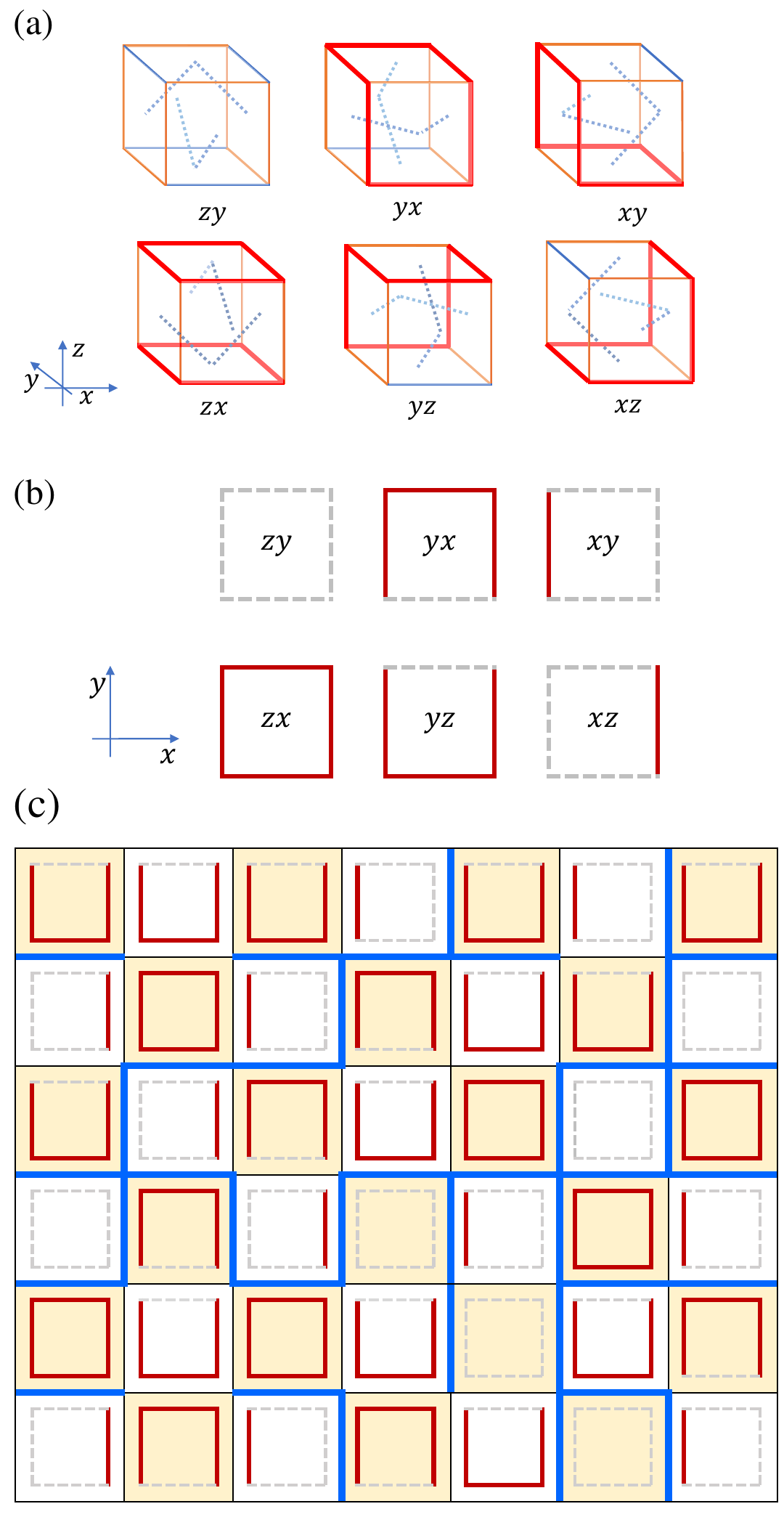}
\caption{\textbf{Realizing defect configurations with Block~C6.} (a) The six orientations of Block~C6. The thick red lines mark the defected edges that occur when replacing one block in an `all $zy$' configuration with each of the other orientations. (b) Top view of the six cases. (c) Checkerboard coloring of the $xy$ plane. Blue edges are needy; when one or three such appear at a vertex, that means that the edge in the negative $z$ direction is also needy.}
\label{fig:C6_defect}
\end{figure}

\rs{Below the top layer, we first fill the entire lattice with blocks oriented $zy$, as shown in Fig.~\ref{fig:C6_defect}a. This `reference' metamaterial has all its defects inside and at the bottom boundary of the top $z$ layer. It realizes some of the defects that we want, but most of them it does not, furthermore it contains defected edges that we do not desire. So there are many edges whose defectedness needs to be changed; we will call these needy. The set of needy edges always satisfies the local parity rule. Now, when the actual orientation of a block is decided, we can think that it is changed (or not) from $zy$ to its final state, and track the corresponding changes in the defects. The possibilities are shown in Fig.~\ref{fig:C6_defect}a. The thick red edges there are those that change from strut to hinge, or vice versa, when $zy$ is replaced with some other orientation.}

\rs{Reorienting a block in a certain $z$ layer amounts to choosing, for its top face --- namely for the square in the $xy$ plane at the interface between this layer and the layer above it in the $z$ direction --- one of the six options shown in Fig.~\ref{fig:C6_defect}b, where the thick red lines mark the change of defectedness of an edge due to reorienting the block from the $zy$ direction to the listed direction.}

\rs{We mark needy edges in blue in the example shown in Fig.~\ref{fig:C6_defect}c. To resolve the neediness of all edges in this $xy$ plane, we require each needy edge to have exactly one thick red line next to it, and each non-needy edge to have zero or two such lines. To achieve this, we introduce the yellow-white checkerboard coloring shown in Fig.~\ref{fig:C6_defect}c. We start with the top $y$ row in this plane, which is certainly such that along it,
\begin{equation}
\label{property}
\tag{*}\text{the upper side of any white square is not needy.}
\end{equation}
The trick will be to maintain this property for all subsequent rows, too. (Here, `top' and `upper' are meant in the sense of the $y$ coordinate.) Note that as we proceed from top to bottom, the set of needy edges will be continuously updated, and \eqref{property} is meant with respect to the set of needy edges just before we start working on a new row.}

\rs{Indeed as long as \eqref{property} holds, it makes sense to carry out the following steps. For each yellow square along our row, we check which of its horizontal sides --- the edges in the $x$ direction --- are needy. We assign the orientation $zy$ if the answer is none; $zx$ if the answer is both; $yx$ if it is only the top; and $yz$ if it is only the bottom. Then, we look at the white squares along the same row and consider their vertical sides --- the edges in the $y$ direction. If none is needy, we assign orientation $zy$ for the block which is responsible for this white square; if both are needy, then $yz$; if only the left or the right side is needy, then $xy$ or $xz$, respectively. This takes care of vertical edges along this row. Furthermore, because of \eqref{property}, needy edges do not remain along the upper boundary, either. }

\rs{Finally, as to the lower boundary, the construction ensures that only the bottom edges of white squares may remain (or become) needy. In terms of the next row, these are top edges of yellow squares. In other words, the upper sides of the white squares of the next row are not needy, that is to say \eqref{property} holds for the next row of squares and the iteration may continue unabated in the negative $y$ direction. After that, there is no impediment to continuing the process in the negative $z$ direction, either, until the construction of the metamaterial is complete.}

\section{Defects realization program}
\label{sec:code}

We consider the following computational challenge: given a particular candidate defect set, our goal is to either find a metamaterial realizing that defect set, or prove that no such metamaterial exists. In the latter case, this implies that no choice of orientations of the blocks generates the specified defect set. Since the number of potential orientations scales exponentially with the number of blocks, direct enumeration of all possibilities is impractical. Furthermore, it is impossible to construct the metamaterial gradually, choosing one orientation at a time, without backtracking when the algorithm later discovers that this leads to a conflict with the defect set somewhere else in the lattice; such backtracking can then lead to an expensive search.

The approach we employed is to map our problem onto an instance of the Boolean satisfiability problem, also known as SAT. This allows us to take advantage of a variety of modern algorithms and codes called ``SAT Solvers'', which are highly efficient for many particular instances of SAT even though exponential scaling cannot always be avoided~\cite{10.1145/3560469}. To use the SAT Solver, we encode the potential orientations of each block in the lattice as Boolean variables. There is a variable for each block and each orientation, with the value ``true'' indicating the block is chosen to be in this orientation, and ``false'' otherwise. Each block can have only one orientation, and each edge induces a constraint over the variables corresponding to the orientations of the adjacent blocks. The constraints are determined by the rules governing the connection between block orientations and edge defectedness, and by whether the defect set specifies a defect for the edge. For example, for Block~C2, a defect on an edge in the $x$-direction generates the constraint that the number of adjacent blocks with orientation in the $x$-direction is even. Once all these constraints are accumulated into a logical statement expressed in the binary variables, the SAT Solver checks whether there exists an assignment of true/false to the variables that satisfies all the constraints. If one is found, it encodes the desired metamaterial; otherwise, the SAT solver returns that no solution exists, which implies that the defect set is unrealizable. Our implementation uses MiniSAT 2.2~\cite{sorensson2010minisat} through PySAT~\cite{imms-sat18}. In our experience, checking realizability of a defect set with Block~C2 on an $L=9$ lattice -- which corresponds to a problem with $9^3=729$ blocks and $3^{729} \approx 10^{347}$ possible metamaterials -- typically takes less than a minute on an ordinary laptop computer.

\section{Concrete example of a knotted defect}
\label{sec:examples}

We describe a construction for Block~C2, based on the grid diagram of Fig.~\ref{fig:grid_example}a, that results in a 3D metamaterial whose defect is a single trefoil knot. This is meant to illustrate the method that we outlined in Sec.~\ref{sec:defects}.

For Block~C2, one of the three principal axes is obviously distinguished. With regard to the standard 3D coordinate system, we use the symbols $x$, $y$, and $z$ to indicate the direction of the distinguished axis; this describes the orientation of each block completely. In fact, it suffices to use a $6\times6\times3$ array, in which we orient the blocks as listed below, 
\emph{from bottom to top}:
\[
\begin{array}{cccccc}
y & y & y & y & y & y\\
y & y & x & x & x & y\\
y & x & y & x & x & y\\
y & x & x & y & x & y\\
y & x & x & x & y & y\\
y & y & y & y & y & y
\end{array};\quad
\begin{array}{cccccc}
y & y & y & y & y & y\\
y & y & z & z & z & y\\
y & z & y & z & z & y\\
y & z & z & y & z & y\\
y & z & z & z & y & y\\
y & y & y & y & y & y
\end{array};\quad
\begin{array}{cccccc}
y & y & y & y & y & y\\
y & y & y & y & y & y\\
y & y & y & y & y & y\\
y & y & y & y & y & y\\
y & y & y & y & y & y\\
y & y & y & y & y & y
\end{array}.
\]
This produces the configuration shown in Fig.~\ref{fig:grid_example}c.

\bibliography{holography}

\end{document}